\documentclass[twocolumn]{autart}    

\usepackage{amsmath}
\usepackage{algorithm,algpseudocode}
\usepackage{amssymb}
\usepackage{color}
\usepackage{enumerate}
\usepackage{wrapfig}
\usepackage{graphicx}          
\usepackage[dvips]{epsfig}    
\usepackage[round, sort]{natbib}
\usepackage{hyperref}
\hypersetup{
	colorlinks = true,
	citecolor = cyan,
}
\usepackage{url}

\newcounter{MYtempeqncnt}
\newtheorem{lemma}{Lemma}[section]
\newtheorem{theorem}{Theorem}[section]

\newtheorem{remark}{Remark}[section]
\newtheorem{definition}{Definition}[section]

\newtheorem{assumption}{Assumption}[section]

\newtheorem{problem}{Problem}

\allowdisplaybreaks

\begin{document}
		\begin{frontmatter}
			
			\title{Robust Control of Unknown Switched Linear\\ Systems from Noisy Data} 
			
			\thanks[footnoteinfo]{The work was supported 
				in part by the National Natural Science Foundation of China under Grants 61925303, 62173034, 62088101, U22B2058, 
				the China Scholarship Council under Grants 202206030127,
				and in part by the Chongqing Natural Science Foundation under Grant 2021ZX4100027. \emph{(Corresponding author: Gang Wang.)}}
			\thanks[footnoteinfo]{This paper was not presented at any IFAC 
				meeting. }
		
		\author[Bit,Tcic]{Wenjie Liu}\ead{liuwenjie@bit.edu.cn}, 
			\author[Bit,Tcic]{Yifei Li}\ead{liyifei@bit.edu.cn},
		\author[Bit,Tcic]{Jian Sun}\ead{sunjian@bit.edu.cn},            
		\author[Bit,Tcic]{Gang Wang}\ead{gangwang@bit.edu.cn}, 
		\author[Bit,Tongji]{Jie Chen}\ead{chenjie@bit.edu.cn} 
		
		\address[Bit]{National Key Lab of Autonomous Intelligent Unmanned Systems and School of Automation,\\ Beijing Institute of Technology, Beijing 100081, China}  
		\address[Tcic]{Beijing Institute of Technology Chongqing Innovation Center, Chonqing 401120, China}     
		\address[Tongji]{Department of Control Science and Engineering, Tongji University, Shanghai 201804, China}        

		\maketitle

		\begin{abstract}
			\indent	This paper investigates the problem of data-driven stabilization for linear discrete-time switched systems with unknown switching dynamics. In the absence of noise, a data-based state feedback stabilizing controller can be obtained by solving a semi-definite program  (SDP) on-the-fly, which automatically adapts to the changes of switching dynamics. However, when noise is present, the persistency of excitation condition based on the closed-loop data may be undermined, rendering the SDP infeasible. To address this issue, an auxiliary function-based switching control law is proposed, which only requires intermittent SDP solutions when its feasibility is guaranteed. By analyzing the relationship between the controller and the system switching times, it is shown that the proposed controller guarantees input-to-state practical stability (ISpS) of the closed-loop switched linear system, provided that the noise is bounded and the dynamics switches slowly enough. Two numerical examples are presented to verify the effectiveness of the proposed controller.
		\end{abstract}

	\end{frontmatter}

	\section{Introduction}\label{sec:intro}
	Since the early nineteenth century, classical control theory has yielded abundant results in areas ranging from feedback control to optimal, adaptive, robust, and nonlinear control. A critical intermediate step for synthesizing a controller and associated stability analysis is acquiring the system model using first-principles or the system identification method. However, with the increasing complexity and inter-connectivity of engineered cyber-physical systems, obtaining system models using first-principles methods has become challenging, while system identification has become data-inefficient and/or computationally expensive. In recent years, thanks to advances in data science and big data technology, data-driven control has emerged as a promising paradigm relative to classical model-based control. Data-driven control methods circumvent the need for parametric system modeling \cite{Chua2018deep} and mitigate the over-fitting of noise \cite{krishnan2021On}, which have attracted significant attention.

	A range of data-driven control methods have been proposed, including 
	iterative feedback tuning \cite{hjalmarsson1998iterative}, adaptive control \cite{aastrom2013adaptive,wu2023data}, and reinforcement learning-based control \cite{sassano2020combining}. Further references can be found in \cite{Hou2013from,persis2020data}. Recently, renewed interest has been generated by the fundamental lemma developed in \cite{willems2005note}, which provides a sufficient condition for the existence of a data-based system representation based on input-state data. This has led to a growing number of publications, including data-enabled predictive control (DeePC) \cite{Coulson2019cdc} and various system analysis and controller design results that use data directly; see \cite{persis2020data} for a recent survey of these developments. The DeePC framework \cite{Coulson2019cdc} aims to design control inputs by solving a convex optimization problem based on system trajectories collected offline. Several extensions have been made, including establishing theoretical stability guarantees in \cite{berberich2019data}, enhancing robustness in \cite{Coulson2021Bridging}, and improving resiliency against attacks in \cite{Liu2021data}. Using input-state data, the work of \cite{persis2020data,van2020data} provides simple data-based parametrizations of linear state feedback systems. Moreover, it has been demonstrated in \cite{persis2020data} that many control problems can be formulated as data-dependent linear matrix inequalities (LMIs), including linear quadratic regulation (LQR) \cite{depersis2020Low,zhao2023data}, robust control \cite{van2020noisy,li2023data}, event- and self-triggered control \cite{li2022robust,Wang2021data,de2022event}, complex and network systems \cite{baggio2020Data}, time-delay systems \cite{rueda2020data,Wang2021delay}, nonlinear systems \cite{guo2022data,hu2023learning}, and system identification \cite{kang2023minimum}.

	All of the aforementioned results focused on systems that can be fully characterized by a finite set of data, such as linear time-invariant (LTI) systems or special nonlinear systems that can be expressed in a `linear-like' form. In such cases, a stabilizing controller designed using offline data can effectively stabilize the system during online implementation. However, when the complexity of a system cannot be fully captured or approximated by a finite set of data, the controller designed offline may no longer be effective  for the changing dynamics of the system. Switched systems, which are commonly used to model real-world systems such as mobile robots \cite{lee2008uniform}, chemical processes \cite{Mhaskar2005predictive}, and power systems \cite{Cardim2009variable,finite2021wang}, are one such example. Switched systems comprise a set of subsystems that switch between each other according to some switching signal. Since the switching signal can be arbitrary in general, even with just two modes (i.e., subsystems), the sequence of activated modes present in the data collected offline may be entirely different from that appearing in the online operation. Therefore, designing a data-driven state feedback control law for switched systems requires modifications to the offline solutions discussed earlier, to account for the abrupt switching dynamics which cannot be fully characterized by a finite set of offline data. 
	
	Several related contributions on data-based stabilization of switched linear systems can be found in \cite{Rotulo2021Online,eising2022using,bianchi2022data}, with the works \cite{eising2022using} and \cite{bianchi2022data} specifically dealing with noise-free and noisy data, respectively. However, the assumption that sufficient input-state trajectories for each subsystem must be collected may seem impractical and hard to validate in general. The results of \cite{Rotulo2021Online} relaxed this assumption by proposing an online perturbed data-based state feedback controller based on a semi-definite program (SDP) that is updated and solved at each time step. Nevertheless, it is important to note that the work \cite{Rotulo2021Online} assumes the offline data are noiseless, which is often not the case in real-world applications. Furthermore, its controller and stability results do not hold when noise is present. 
	
	
	The aim of this paper is to extend the findings of \cite{Rotulo2021Online} by incorporating noisy data. The challenge in implementing an online controller lies in ensuring that the closed-loop data are persistently exciting, which is crucial for recovering the system behavior and for parameterizing the state feedback controller using solutions from the data-based SDP \cite{depersis2020Low}. To address this challenge, this paper proposes a switched controller that employs a switching control law based on variations of an auxiliary function. A robust data-based SDP is formulated using noisy data collected in real time, which is solved intermittently only when its feasibility is guaranteed. By establishing the relationship between the controller's switching times and the system's switching times, we show that the closed-loop system is ISpS under mild assumptions on the noise and switching dynamics.

	In summary, the paper offers the following contributions to data-based stabilization of unknown linear switched systems.
	\begin{itemize}
		\item [c1)]
		A novel online data-based switched controller is designed, following an auxiliary function based switching law that dictates whether a data-based SDP is solved or not.
		\item [c2)]		
		Conditions on the noise and control inputs are provided to guarantee the feasibility of the robust data-based SDP.
		\item [c3)]
		The relationship between the system switching times and the controller switching times is developed, and ISpS of the closed-loop system is established.
	\end{itemize}

	\emph{Notation:}
	Denote the sets of real numbers, integers, and positive integers by $\mathbb{R}$, $\mathbb{N}$, and $\mathbb{N}_{+}$ respectively.
	For a matrix $M$, if it has full column rank, its left pseudo-inverse is denoted by $M^\dag$.
	Given a vector $x\in \mathbb{R}^{n_x}$, let $\Vert x\Vert$ denote its Euclidean norm.
	Denote  the spectral norm of a matrix $M$ by $\Vert M\Vert$.
	Given a measurable time function $f : \mathbb{N} \rightarrow \mathbb{R}^n$ and a time interval $[0,k)$ we denote the $\mathcal{L}_{\infty}$ norm of $f(\cdot)$ on $[0,k)$ by $\Vert f_k\Vert_{\infty} := {\rm ess}\sup_{s \in [0,k)} \Vert f(s)\Vert$.
	For matrices $A$, $B$, and $C$ with compatible dimensions, we abbreviate $ABC(AB)^\prime$ as $AB \cdot C[\star]^\prime$.
	Let $\underline{\lambda}_{P}$ [$\overline{\lambda}_{P}$] be the minimum (maximum) singular value of matrix $P$.
	Let $x_{[k_1, k_2]} := [x_{k_1}~x_{k_1 + 1}~ \cdots~ x_{k_2}]$ denote a stacked window of signal $x$ in discrete time interval $[k_1, k_2]$. 
	The Hankel matrix associated with sequence $\{x(k)\}_{k = 0}^{N - 1}$ is denoted by
	\begin{equation*}
		H_{L}(x_{[0,N-1]}):=\left[
		\begin{matrix}
			x_{0} & x_{1} & \ldots & x_{N-L} \\
			x_{1} & x_{2} & \ldots & x_{N-L+1} \\
			\vdots & \vdots & \ddots & \vdots \\
			x_{L-1} & x_{L} & \ldots & x_{N-1}
		\end{matrix}
		\right].
	\end{equation*}
	The definition of  $\mu$-persistent excitation as in \cite[Definition 3.1]{coulson2022robust} is given below.
	\begin{definition}[{$\mu$-persistently exciting}]\label{def:pe}
		Let $\mu>0$.
		A signal $\{x_t\}_{t = 0}^{N - 1} \in \mathbb{R}^{n}$ with $N \ge (n + 1)L + 1$ is $\mu$-persistently exciting of order $L$ if $\underline{\lambda}_{H_{L}(x_{[0,N - 1]})}\ge \mu$.
	\end{definition}
	
	
\section{Preliminaries and Problem Formulation}
\label{sec:preliminaries}

In this section, we begin by reviewing the results in \cite{depersis2020Low}, which dealt with the data-based stabilization of unknown linear time-invariant (LTI) systems using noisy data.
This plays an instrumental role in deriving and explaining our results.

\subsection{Data-driven control of LTI systems}\label{sec:preliminaries:sdp}

	
	Consider a discrete-time LTI system as follows
	\begin{equation}\label{eq:sys_lti_noise}
		x(k + 1) = Ax(k) + Bu(k) + d(k), \quad k \in \mathbb{N}
	\end{equation}
	where $x(k) \in \mathbb{R}^{n_x}$ is the state, $u(k) \in \mathbb{R}^{n_u}$ is the control input, and $d(k) \in \mathbb{R}^{n_x}$ is the noise or disturbance.
	The system matrices $(A, B)$ are unknown and we do not have access to the disturbance $d(k)$. Instead, we assume there are some input-state data $(u_{[-T, -1]},x_{[-T, 0]})$ obtained from e.g.,  offline experiments by exciting the system using control inputs  $u_{[-T, -1]}$ and collecting the corresponding states $x_{[-T, 0]}$.
	For consistency, negative indices are used to refer to data collected offline.
	Define the data matrices as follows 
	\begin{subequations}\label{eq:X0}
		\begin{align}
			U_{-1} &= [u(-T)~u(-T +1)~\cdots~u(- 1)],\\
			X_{-1} &= [x(-T)~x(-T +1)~\cdots~x(- 1)],\\
			X_{0} &= [x(-T +1)~x(-T +2)~\cdots~x(0)],\\
			D_{-1} &= [d(-T)~d(-T+1)~\cdots~d(- 1)],\\
			 W_{-1} &= \left[U_{-1}^{\prime}~X_{-1}^{\prime}\right]^{\prime}\label{eq:X0:W}.
		\end{align}
	\end{subequations}
	
	A method for finding a matrix $K$ such that $A + BK$ is Schur stable using disturbance-corrupted data in \eqref{eq:X0} was presented in \cite[Theorem 4]{depersis2020Low}.
	\begin{lemma}\label{lem:sdp_lti_noise}
		Let $U_{-1}$, $X_{-1}$ and $X_{0}$ be data generated by system \eqref{eq:sys}.
		If the condition
		\begin{equation}\label{eq:rankW0}
			{\rm rank}(W_{-1}) = n_x + n_u
		\end{equation}
		holds, then there exists a constant $\delta >0$ such that for $\Vert D_{-1}\Vert \le \delta$ the following problem is feasible
		\begin{align}\label{eq:sdp_lti_noise}
			&\min_{(\gamma, Q, P, L, V)} \gamma\nonumber\\
			&{\rm subject~to}\nonumber\\
			&\left\{
			\begin{aligned}
				&X_0Q P^{-1}Q^\prime X_0^\prime - P+ I \preceq 0 \\
				&P \succeq I\\
				&L - U_{-1}QP^{-1}Q^\prime U_{-1}^{\prime} \succeq 0\\
				&V - QP^{-1}Q^\prime \succeq 0\\
				&X_{0}Q = P\\
				&{\rm tr}(P) + {\rm tr}(L) + \alpha {\rm tr}(V)\preceq \gamma	
			\end{aligned}
			\right.
		\end{align} 
		where $\alpha > 0$ is arbitrary. 
		Let $(\bar{\gamma}, \bar{Q},\bar{P},\bar{L},\bar{V})$ be an optimal solution of \eqref{eq:sdp_lti_noise}. 
		Then the matrix $K^* = U_{-1} \bar{Q} \bar{P}^{-1}$ is stabilizing.
	\end{lemma}

	In light of condition \eqref{eq:rankW0}, it is notable that a stabilizing gain $K^*$ can be parameterized using solely data. This implies that the state feedback controller $u(k) = K^*x(k)$ can be implemented on system \eqref{eq:sys_lti_noise}. According to \cite[Theorem 3.1]{coulson2022robust}, for any bounded disturbance, i.e., $\Vert D_{-1}\Vert \le \delta$ for some $\delta>0$, the condition \eqref{eq:rankW0} can be ensured by utilizing a sufficiently exciting input sequence.
	
	We present Lemma \ref{lem:wpe} below, which shows that a $\bar{w}$-persistently exciting input sequence  of order $n_x + 1$ is sufficient to satisfy the condition \eqref{eq:rankW0}. We postpone the proof of this lemma to Appendix \ref{pf:lem:wpe}.
	
	\begin{lemma}
		\label{lem:wpe}
		Suppose the system \eqref{eq:sys_lti_noise} is controllable. For any $\delta \ge 0$, let the disturbance obey $\Vert D_{-1}\Vert \le \delta$. Then there exists a constant $\bar{w} >0$ such that if the input sequence ${u(-T),\cdots,u(-1)}$ is $\bar{w}$-persistently exciting of order $n_x + 1$, then the condition \eqref{eq:rankW0} is satisfied.
	\end{lemma}
	
	\subsection{Switched system and problem formulation}\label{sec:preliminaries:sys}
	Consider the following discrete-time linear switched system 
	\begin{equation}\label{eq:sys}
		x(k + 1) = A_{\sigma(k)}x(k) + B_{\sigma(k)}u(k) + d(k),
	\end{equation}
	where the switching signal $\sigma : \mathbb{N} \rightarrow \mathcal{M}$ is a piecewise constant function of time taking values in the finite set $\mathcal{M}:= \{1, 2,$ $ \cdots, m\}$, where $m > 1$ is the total number of modes.	The matrices $(A_{\sigma(k)}, B_{\sigma(k)})$ belong to a collection of constant matrices $ \mathcal{S} := \{(A_i, B_i):i \in \mathcal{M}\}$.	Let  $k_{s_j}$ denote the time when the $j$-th switching occurs, i.e., $k_{s_j} = \min\{k >k_{s_{j - 1}}:\sigma(k) \ne \sigma(k_{s_{j - 1}}) \}$ with $j \in \mathbb{N}_{+}$. Without loss of generality, let $k_{s_0}=0$.	Suppose that the active mode selected by $k_{s_j}$ is indicated by $i$, so it holds that $i = \sigma(k)$ for all $k \in [k_{s_j}, k_{s_{j + 1}} - 1]$.
	
	In this paper, we make the following assumptions.
	\begin{assumption}
		[Unknown system]\label{as:sys}
		The pairs $(A_i, B_i)$ for all $i \in \mathcal{M}$, the switching signal $\sigma$, and the switching instant $k_{s_j}$ with $j \in \mathbb{N}_+$ are unknown.
	\end{assumption}
	\begin{assumption}[{Controllability}]\label{as:ctrl}
		For each $i \in \mathcal{M}$, the pair $(A_{i}, B_{i})$ is controllable.
	\end{assumption}
	\begin{assumption}
		[Bounded disturbance]
		\label{as:noise}
		For all $k \in \mathbb{N}$, 	it holds that	 $ d(k) \in \mathbb{B}_{\bar{d}} := \{d|\Vert d\Vert \le \bar{d}\}$ for some known constant $\bar{d} \ge 0$. 
	\end{assumption}
	
	In addition, let us suppose that the controller side possesses a buffer of size $T \in \mathbb{N}_{+}$. At each time instant $k \in \mathbb{N}$, the buffer records the latest $T$ input-state samples, which are collected in $U_{k - 1} $, $X_{k - 1} $, and $X_{k } $, as follows
	\begin{subequations}\label{eq:Xk}
		\begin{align}
			U_{k - 1} &= [u(k - T)~u(k - T + 1)~\cdots~u(k - 1)],\\
			X_{k - 1}& = [x(k - T)~x(k - T + 1)~\cdots~x(k - 1)],\\
			X_{k} &= [x(k - T + 1)~x(k - T + 2)~\cdots~x(k)].
		\end{align}
	\end{subequations}
	
	Let  $D_{k - 1} = [d(k - T)~d(k - T + 1)~\cdots~d(k - 1)]$ denote the disturbance matrix corresponding to the most recent $T$ input-state samples in $U_{k-1}$, $X_{k-1}$, and $X_k$. Notice that some $k\in[0,T-1]$, the indices of the samples in \eqref{eq:Xk} are negative, which refers to data obtained offline, as described in Section \ref{sec:preliminaries:sdp}. Specifically, there exists a positive constant $\bar{w} > 0$ such that the input-state data $[U_{-1}^{\prime} X_{-1}^\prime]'$ generated from the system \eqref{eq:sys} by using a $\bar{w}$-persistently exciting input sequence ${u(-T), \cdots, u(-1) }$ of order $n_x + 1$, has full row rank, according to Lemma \ref{lem:wpe}. To simplify the analysis, we assume that the samples in $X_{-1}$, $X_{0}$ are generated from the same subsystem, which implies that the SDP \eqref{eq:sdp_lti_noise} is feasible with $X_{-1}$, $X_0$, and $U_{-1}$. 
	At each $k\ge T$, the buffer's window is shifted one step forward, which means that the oldest sample (i.e., the first column of the data matrices in \eqref{eq:Xk}) is removed, and the new sample is added to the buffer.


	In order to stabilize the system \eqref{eq:sys} under the presence of disturbances and unknown switching modes, we aim to design a control signal $u(k)$ for the aforementioned setups. To accomplish this, we propose an online switched controller composed of an exciting signal $\epsilon(k)$ and a dynamic state feedback law $K(k)$, as given by
	\begin{equation}\label{eq:isps:uk}
		u(k) \!= \!
		\begin{cases}
			\epsilon(k), \!&{\rm if}~k \in  \mathbb{I}_{E}
			\\
			K(k)x(k),\!&{\rm else}
		\end{cases}
	\end{equation}
	Here, the exciting signal $\epsilon(k)$ is appropriately selected from the set $\epsilon(k) \in \mathcal{B}_{\delta_{\epsilon}} := \{\epsilon\,|\,\Vert \epsilon(k)\Vert \le \delta_{\epsilon}\}$ to ensure the persistency of excitation of the noisy input-state data sequence collected online. The set $\mathbb{I}_{E}$ is a collection of some event times governed by a switching law, and $K(k) \in \mathbb{R}^{n_u \times n_x}$ is a dynamic state feedback control gain. The design of $\epsilon(k)$, $K(k)$, and $\mathbb{I}_{E}$ will be discussed later.

	
	Moreover, to reflect the goal of stabilization under the unknown disturbance $d(k)$ and the exciting signal $\epsilon(k)$, we invoke the input-to-state practical stability (ISpS).
	The definition of ISpS treating $d(k)$ as an unknown input is adapted from Definition 2.2 in \cite{Z1994Small}.
	
	\begin{definition}[ISpS \cite{Z1994Small}]\label{def:isps}
		System \eqref{eq:sys} in closed-loop with a control signal as in \eqref{eq:isps:uk} is ISpS if, for any $x(0) \in \mathbb{R}^{n_x}$ and measurable essentially bounded $d(k)$ on $k \in [0,+\infty)$, its solution satisfies
		\begin{equation}\label{eq:isps}
			\Vert x(k)\Vert \le \alpha(\Vert x(0)\Vert, k) + \beta(\Vert d_k\Vert_{\infty}) + c_0,\quad \forall k \in \mathbb{N} 
		\end{equation}
		where $\alpha$ is a $\mathcal{KL}$-function, $\beta$ is a $\mathcal{K}$-function, and $c_0$ is some constant\footnote{
			A function $\beta : [0,\infty) \rightarrow [0, \infty)$ is said to be of class $\mathcal{K}$ if it is continuous, strictly increasing, and $\beta(0) = 0$.
			A function $\alpha : [0,\infty)\times [0,\infty) \rightarrow [0, \infty)$ is a $\mathcal{KL}$-function if $\alpha(\cdot, k)$ is of class $\mathcal{K}$ for each fixed $k \ge 0$ and $\alpha(s, k)$ decreases to $0$ as $k \rightarrow \infty$ for any fixed $s \in \mathbb{N}$.}.
\end{definition}

With the preliminaries above, the problem to be addressed is formally stated as follows.
\begin{problem} \label{pro:isps}
	For the switched system \eqref{eq:sys} under Assumptions \ref{as:sys}--\ref{as:noise}, design a data-based control input of the form in \eqref{eq:isps:uk} to ensure ISpS of the closed-loop system.
\end{problem}

\section{Robust Data-driven Switched Control}
\label{sec:ddcontrol_strategy}

To tackle Problem \ref{pro:isps}, we draw inspiration from the approach presented in \cite{Rotulo2021Online}. In this approach, the state feedback control gain $K(k)$ is constructed based on an online version of the SDP \eqref{eq:sdp_lti_noise}, which allows for automatic adaptation of the control input to the switching dynamics. Specifically, we formulate a robust SDP at each $k \in \mathbb{N}$ using the noisy data $U_{k-1}$, $X_{k-1}$, and $X_k$, as follows
\begin{align}\label{eq:sdp_noise}
&\min_{(\gamma, Q,P,L, V)} \gamma\nonumber\\
&{\rm subject~to}\nonumber\\
&\left\{
\begin{aligned}
	&X_{k}QP^{-1}Q^\prime X_{k}^\prime - P + I \preceq 0 \\
	&P \succeq I\\
	&L - U_{k - 1}QP^{-1}Q^\prime U_{k - 1}^{\prime} \succeq 0\\
	&V - QP^{-1}Q^\prime \succeq 0\\
	&X_{k - 1}Q = P\\
	&{\rm tr}(P) + {\rm tr}(L) + \alpha {\rm tr}(V)\preceq \gamma	
\end{aligned}
\right.
\end{align}
Here, $\alpha >0$ is chosen arbitrarily to balance performance and robustness, as in Lemma \ref{lem:sdp_lti_noise}. 
Suppose that SDP \eqref{eq:sdp_noise} is feasible at time $k$, and we denote its optimal solution by $(\gamma^*(k), Q^*(k),P^*(k), L^*(k), V^*(k))$.
An important condition for the feasibility of SDP \eqref{eq:sdp_noise} at time $k$ is the rank condition:
\begin{equation}\label{eq:rankWk}
{\rm rank}(W_{k - 1}) = \left[
\begin{matrix}
	U_{k - 1}\\
	X_{k - 1}
\end{matrix}
\right] = n_u + n_x
\end{equation}
where $n_u$ and $n_x$ are the dimensions of the input and state vectors, respectively. 

According to Lemma \ref{lem:wpe}, this condition can be satisfied if the input sequence ${u(k - T),\cdots,u(k - 1)}$ is $\bar{w}$-persistently exciting of order $n_x + 1$ for some $\bar{w} >0$. Note from \eqref{eq:isps:uk} that if $u(i) = \epsilon(i)$ for all $i \in [k - T, k - 1]$, the persistency of excitation of the input sequence can be easily guaranteed.
However, using $u(k) = \epsilon(k)$ places the system \eqref{eq:sys} in open-loop, which can harm system stability. Moreover, the SDP \eqref{eq:sdp_noise} may not be feasible when matrices $X_{k - 1}$ and $X_{k}$ contain data generated from different subsystems. Even if it is feasible, the resulting matrix $K(k)$ constructed from its optimal solution may not be stabilizing. 
{This indicates that there are times when solving SDP \eqref{eq:sdp_noise} is unnecessary. In other words, using the open-loop control signal $\epsilon(k)$ sparingly may have little influence on stability at times.}

In pursuit of addressing Problem \ref{pro:isps}, we are faced with three fundamental questions: i) How can we ensure the feasibility of SDP \eqref{eq:sdp_noise}? ii) When should we solve SDP \eqref{eq:sdp_noise} and when can we use $\epsilon(k)$ instead? iii) What about the ISpS of the closed-loop system? In the upcoming sections, we provide a comprehensive explanation of the designated controller in \eqref{eq:isps:uk} and answer each of these questions.

Throughout this paper, we operate under the following two assumptions.
\begin{assumption}[{Data length}]\label{as:T}
The number of samples in \eqref{eq:Xk} satisfies $T \ge 2N - 1$, where $N = (n_x + 1)n_u + n_x$ is necessary for the persistency of excitation of order $n_x + 1$.
\end{assumption}
\begin{assumption}[{Dwell time}]\label{as:dwell}
The dwell time $\tau := \min_{j \in \mathbb{N}}~k_{s_{j + 1}} - k_{s_j}$ satisfies $\tau > T$.
\end{assumption}

It is important to note that Assumption \ref{as:T} has been utilized in \cite{Rotulo2021Online} to ensure the feasibility of the SDP \eqref{eq:sdp_noise} in the disturbance-free case. However, as the complexity of solving \eqref{eq:sdp_noise} scales with $T$, smaller values of $T$ are often preferred for implementation purposes. Without loss of generality, we set $T := 2N - 1$ in the subsequent analysis.

Moreover, Assumptions \ref{as:T} and \ref{as:dwell} guarantee that the system switches slowly enough such that the collected input-state data are generated by at most two subsystems, and at least $N$ input-state data are generated from the same subsystem. This simplifies the analysis to the question of ensuring the feasibility of the SDP \eqref{eq:sdp_noise}, which we address in the following subsections.



\subsection{Robust data-driven controller}

%

The matrix $K(k)$, the signal $\epsilon(k)$, and the switching laws in $\mathbb{I}_E$ determine the activated times of signal $\epsilon(k)$ in \eqref{eq:isps:uk}, as specified below. To decide when to solve SDP \eqref{eq:sdp_noise} and when to apply $\epsilon(k)$, an auxiliary function $\mathcal{V}(x(k)) = x'(k) P(k) x(k)$ is designed in conjunction with the controller in \eqref{eq:isps:uk}. 
Specifically, for $k = 0$, as mentioned in Section \ref{sec:preliminaries:sys}, ${\rm rank}(W_{-1}) = n_x + n_u$, and samples in matrices $X_{-1}$, $X_0$ are generated from the same subsystem. Therefore, SDP \eqref{eq:sdp_noise} is feasible, as per Lemma \ref{lem:sdp_lti_noise}. Let $P^*(0)$ be an optimal solution of SDP \eqref{eq:sdp_noise} at $k=0$. The initial condition is given by $P(0) = P^*(0)$. For $k \in \mathbb{N}_+$, matrices $P(k)$ are the $P$-solutions of SDP \eqref{eq:sdp_noise}, which will be specified later. The confidence about the feasibility of SDP \eqref{eq:sdp_noise} is measured by function $\mathcal{V}(x(k))$.

We begin by introducing some definitions before proceeding.  
Fix any small constant $\delta_{V} >0$, and without loss of generality, assume that the initial condition $x(0) \notin \ker(P(0))$ and $\mathcal{V}(x(0)) >\delta_V$. The set of times $k\in\mathbb{N}$ such that $\mathcal{V}(x(k)) \le \delta_V$ is denoted by $ \mathbb{I}_{\delta_V}:=\{k \in \mathbb{N}\,|\, \mathcal{V}(x(k)) \le \delta_V\}$. For some $\lambda_0 \in (0,1)$ and all $k \notin \mathbb{I}_{\delta_V}$, let $\{k_j \}_{j \in \mathbb{N}}$ be the times $k_j$ when $\mathcal{V}(x(k_j)) > \lambda_0 \mathcal{V}(x(k_j - 1))$ and $\mathcal{V}(x(k_j - 1)) \le \lambda_0 \mathcal{V}(x(k_j - 2))$. Set $k_0 = 0$. Similarly, let $\{k^j \}_{j \in \mathbb{N}}$ be the times $k^j$ when $\mathcal{V}(x(k^j - 1)) > \lambda_0 \mathcal{V}(x(k^j - 2))$ and $\mathcal{V}(x(k^j)) \le \lambda_0 \mathcal{V}(x(k^j - 1))$. Assume that the elements in $\{k_j \}_{j \in \mathbb{N}}$ and $\{k^j \}_{j \in \mathbb{N}}$ are ordered chronologically, adhering to $0= k_0 < k_0 + N < k^0 < k_1 < k_1 + N <k^1<\cdots$. This ordering will be proved in the next subsection.

Expanding on the definition above, we provide a specification for the controller \eqref{eq:isps:uk} as follows
\begin{equation}\label{eq:uk}
u(k) \!= \!
\begin{cases}
\epsilon(k), \!\!&\!\!{\rm if}~k \in [k_{j}, k_{j} + N - 1]\\
K(k)x(k),\!\!&\!\!{\rm else}
\end{cases}
\end{equation}
where $\epsilon(k)$ is chosen within the ball $\mathcal{B}_{\delta_{\epsilon}} := \{\epsilon\,|\,\Vert \epsilon(k)\Vert \le \delta_{\epsilon}\}$, and the controller gain $K(k)$ is set to
\begin{equation}\label{eq:kstrategy}
K(k) = U_{k - 1}Q(k)P(k)^{-1}.
\end{equation}
Furthermore, $P(k)$ and $Q(k)$ are determined by 
\begin{align}\label{eq:P_strategy}
&(P(k),Q(k)) = \nonumber\\
&\qquad \begin{cases}
(P^*(k),Q^*(k))~ {\rm from}~\eqref{eq:sdp_noise},\!\!\!&{\rm if}~ k \in [k_j + N, k^j] \\
(P(k - 1),Q(k - 1)), \!\!\!&{\rm else}.
\end{cases}
\end{align}

To implement the controller given by \eqref{eq:uk}--\eqref{eq:P_strategy}, we must solve the SDP given by \eqref{eq:sdp_noise} for all $k \in [k_j + N, k^j]$ for each $j\in \mathbb{N}$. The feasibility of this SDP is established in the next subsection.

\subsection{Feasibility of SDP \eqref{eq:sdp_noise}}\label{sec:onlinecontrol:feasibility_SDP}

Based on the analysis presented in the previous section, the feasibility of SDP \eqref{eq:sdp_noise} at time $k$ depends on the rank condition \eqref{eq:rankWk} and the samples in data matrices $X_{k-1}$ and $X_k$. Note that for $k \in [k_{s_j}+1,k_{s_j}+T-1]$, the data matrices contain a mixture of samples from subsystems $\sigma(k_{s_j})$ and $\sigma(k_{s_{j-1}})$, as indicated by Assumptions \ref{as:T} and \ref{as:dwell}. Therefore, even if condition \eqref{eq:rankWk} is satisfied, SDP \eqref{eq:sdp_noise} may be infeasible.

To address this issue, this subsection conducts the feasibility analysis of SDP \eqref{eq:sdp_noise} in two steps. First, we show that if condition \eqref{eq:rankWk} is satisfied at time $k$, then SDP \eqref{eq:sdp_noise} is always feasible, provided that $\bar{d}$ is small enough. 
Moreover, if $k \in [k_{s_j} + T, k_{s_{j+1}}]$, then the resultant matrix $K(k)$ in \eqref{eq:kstrategy} is stabilizing. Second, we derive conditions on $\epsilon(k)$ and $\bar{d}$ that guarantee \eqref{eq:rankWk} always holds at times $k \in [k_j+N,k_j]$, thus confirming the feasibility of the controller \eqref{eq:uk}--\eqref{eq:P_strategy}.

In light of Assumptions \ref{as:T} and \ref{as:dwell}, the time interval $[k_{s_j}+1, k_{s_{j+1}}]$ is partitioned into two sub-intervals: $[k_{s_j}+1, k_{s_j}+T-1]$ and $[k_{s_j}+T, k_{s_{j+1}}]$. For $k\in[k_{s_j}+1, k_{s_j}+T-1]$, the data matrices $X_{k-1}$ and $X_k$ comprise a mixture of states from subsystems $\sigma(k_{s_j})$ and $\sigma(k_{s_{j-1}})$. In contrast, data matrices in the second sub-interval collect only samples generated from subsystem $\sigma(k_{s_j})$. The feasibility analysis of SDP \eqref{eq:sdp_noise} proceeds by examining these two sub-intervals.

Before stating the main results, the disturbance-free version of SDP \eqref{eq:sdp_noise}, which was used in \cite{Rotulo2021Online}, is introduced below:
\begin{align}\label{eq:sdp_ideal}
&\min_{(\gamma, Q,P,L)} \gamma\nonumber\\
&{\rm subject~ to}\nonumber\\
&\left\{
\begin{aligned}
&(X_{k} \!-\! D_{k - 1})QP^{-1}Q^\prime(X_k \!-\! D_{k -1})^\prime \!-\! P \!+\! I \preceq 0 \\
&P \succeq I\\
&L - U_{k - 1}QP^{-1}Q^\prime U_{k - 1}^{\prime} \succeq 0\\
&X_{k - 1}Q = P\\
&{\rm tr}(P) + {\rm tr}(L) \preceq \gamma.
\end{aligned}
\right.
\end{align}	

The optimal solutions of this SDP above will be utilized to construct feasible solutions for SDP \eqref{eq:sdp_noise}.

With this in mind, the following lemmas demonstrate the feasibility of SDP \eqref{eq:sdp_noise} for $k\in[k_{s_j}+1, k_{s_j}+T-1]$ and $k\in[k_{s_j}+T, k_{s_{j+1}}]$, respectively. Their proofs can be found in Appendices \ref{sec:lem:feasible_1} and \ref{sec:lem:feasible_2}.

\begin{lemma}\label{lem:feasible_1}
Under Assumptions \ref{as:sys}--\ref{as:dwell}, let $X_k$, $X_{k-1}$, and $U_{k-1}$ be collected from system \eqref{eq:sys}. Let $i \in \mathcal{M}$ denote the subsystem selected by $\sigma(k_{s_j})$, i.e., $\sigma(k_{s_j}) = i$. Consider SDP \eqref{eq:sdp_noise} with any $\alpha > 0$. For every constant $\bar{w} \geq 0$, there exists a constant $\delta_{d,1} \geq 0$ such that if the following conditions hold: i) the input sequence ${u(k-T),\ldots,u(k-1)}$ is $\bar{w}$-persistently exciting of order $n_x+1$, ii) $k\in[k_{s_j}+T,k_{s_{j+1}}]$, and iii) $\bar{d}<\delta_{d,1}$, then condition \eqref{eq:rankWk} holds and SDP \eqref{eq:sdp_noise} is feasible. Let  $(\gamma^*(k), Q^*(k),P^*(k), L^*(k), V^*(k))$ be an optimal solution of SDP \eqref{eq:sdp_noise}. The matrix $K^*(k) = U_{k - 1}Q_{i}^*(k)P_{i}^*(k)^{-1}$ is such that $A_i + B_i K^*(k)$ is Schur stable.
\end{lemma}

\begin{lemma}\label{lem:feasible_2}
Under Assumptions \ref{as:sys}--\ref{as:dwell}, let $X_k$, $X_{k-1}$, and $U_{k-1}$ be collected from system \eqref{eq:sys}. Consider SDP \eqref{eq:sdp_noise} with any $\alpha > 0$. For constants $\bar{w}$ and $\delta_{d,1}$ specified in Lemma \ref{lem:feasible_1}, if the following conditions hold: i) the input sequence ${u(k-T),\ldots,u(k-1)}$ is $\bar{w}$-persistently exciting of order $n_x+1$, ii) $k\in[k_{s_j}+1,k_{s_j}+T-1]$, and iii) $\bar{d}<\delta_{d,1}$, then condition \eqref{eq:rankWk} holds and SDP \eqref{eq:sdp_noise} is feasible.
\end{lemma}

Lemmas \ref{lem:feasible_1} and \ref{lem:feasible_2} are based on the assumption that the input sequence ${u(k - T),\cdots, u(k -1) }$ is $\bar{w}$-persistently exciting of order $n_x + 1$ at time instant $k$.
The following two lemmas present conditions on $\epsilon(k)$ and $\bar{d}$, which ensure that the persistency of excitation holds for all $k \in [k_j + N, k^j]$ with $j \in \mathbb{N}$. The proofs are provided in Appendices \ref{sec:app:rank} and \ref{sec:app:lem:relationship}.

\begin{lemma}\label{lem:rank}
For any $j\in \mathbb{N}$ and $\delta_{\epsilon} >0$, there exists a constant $\bar{w}>0$ such that the sequence ${\epsilon(k_j),\cdots, \epsilon(k_j + N - 1)}$ can be designed to be $\bar{w}$-persistently exciting of order $n_x + 1$, with $\epsilon(k) \in \mathbb{B}_{\delta_\epsilon}$ for all $k \in [k_j, k_j + N-1]$.
If $\bar{d} < \delta_{d,1}$, where $\delta_{d, 1}$ is given in Lemma \ref{lem:feasible_1}, then the SDP \eqref{eq:sdp_noise} is feasible for all $k\in [k_j + N, k_j +T]$.
\end{lemma}

We can observe from \eqref{eq:P_strategy} that the SDP \eqref{eq:sdp_noise} should be feasible for $k \in [k_j + N, k^j]$. 
In light of Lemma \ref{lem:wpe}, this can be ensured if $k_j +T\ge k^j$. 
To show this, we first construct a set of Lyapunov candidates.
According to Assumption \ref{as:ctrl}, for each subsystem $i \in \mathcal{M}$, there exist constant $\beta_i>0$ and matrices $P_i$, $K_i$ such that
$\mathcal{A}_i^{\prime}P_i\mathcal{A}_i - P_i = -\beta_i I$ holds with $\mathcal{A}_i = A_i + B_i K_i$.
Consider Lyapunov candidate functions $\{\mathcal{W}_i(x(k))= x(k)' P_i x(k)\,|\, i\in \mathcal{M}\}$, each  obeying \begin{equation}\label{eq:boundV}
\underline{\lambda}_{P} \Vert x(k)\Vert^2 \le \mathcal{W}_i(x(k)) \le \bar{\lambda}_{P} \Vert x(k)\Vert^2
\end{equation}
where $\underline{\lambda}_P = \min_{i \in \mathcal{M}} \underline{\lambda}_{P_i}$ and $\bar{\lambda}_{P} = \max_{i \in \mathcal{M}} \bar{\lambda}_{P_i}$ denote the minimum and maximum eigenvalues of $P_i$.
Utilizing the Lyapunov functions above, the following lemma reveals the relationships between $k_j +T$, $k^j$, and the switching time $k_{s_j}$.

\begin{lemma}\label{lem:relationship}
There exist constants $\delta_{d,2}$ and $0<\check{\lambda}_0 <\hat{\lambda}_0\le 1$ such that for all $\bar{d} < \delta_{d,2}$ and $\lambda_0 \in [\check{\lambda}_0, \hat{\lambda}_0)$, the following facts hold for all $j \in \mathbb{N}$: i) $k_j = k_{s_j}$, and ii) $k^j = k_{s_j} + T$.
Moreover, let $i \in\mathcal{M}$ denote the subsystem selected by $\sigma(k_{s_j})$. 
For all $k \in [k_{s_j} + T, k_{s_j + 1}-1]$, it holds that $\mathcal{W}_i(x(k + 1)) \le \lambda_0 \mathcal{W}_i(x(k))$.
\end{lemma}

The above lemmas provide solutions for questions i) to iii). 
For simplicity, we consider the worst case scenario where any matrix $K_i$ that stabilizes subsystem $(A_i, B_i)$ cannot stabilize subsystems $(A_j, B_j)$ for all $j \ne i \in \mathcal{M}$. We will  derive conditions for the ISpS.

\subsection{Stability analysis}

This section investigates the stability of switched system \eqref{eq:sys} under the controller strategy \eqref{eq:uk}--\eqref{eq:P_strategy}.
Noticing from Lemma \ref{lem:feasible_2} that $K(k)$ is not stabilizing for $k \in[k_{s_j}, k_{s_j} + T-1]$, our analysis is carried out in two steps. 
First, we establish the uniform boundedness of matrix $K(k)$ in \eqref{eq:kstrategy} for all $k \notin [k_j, k_j +N - 1]$, where $j \in \mathbb{N}$. 
Using the fact that $\epsilon(k) \in \mathbb{B}_{\delta_\epsilon}$ holds for all $k \in [k_j, k_j +N - 1]$, it can be deduced that the state remains bounded for all $k \in[k_{s_j}, k_{s_j} + T-1]$.
Building on these findings and recalling Lemma \ref{lem:relationship}, we prove the ISpS provided that the disturbance is upper bounded and the dynamics switches slowly enough.

We use Lemmas \ref{lem:feasible_1}--\ref{lem:relationship} to derive an upper bound on $\Vert K(k)\Vert$ as follows.

\begin{theorem}\label{thm:k_bound}
Consider the switched system \eqref{eq:sys} subject to Assumptions \ref{as:sys}--\ref{as:dwell}. If $\bar{d} < \delta_{d,2}$, and $\lambda_0 \in [\check{\lambda}_0, \hat{\lambda}_0)$, where $\delta_{d,2}$, $\check{\lambda}_0$, and $\hat{\lambda}_0$ are as defined in Lemma \ref{lem:relationship}, then there exists a constant $\delta_K > 0$ such that matrix $K(k)$ in \eqref{eq:kstrategy} satisfies $\Vert K(k)\Vert \le \delta_K$ for all $k \notin [k_j, k_j +N - 1]$, where $j \in \mathbb{N}$.
\end{theorem}

\begin{pf}

Suppose that the SDPs \eqref{eq:sdp_noise} and \eqref{eq:sdp_ideal} are feasible at time $k$. The proof of Lemmas \ref{lem:feasible_1} and \ref{lem:feasible_2} has shown that there exists  constant $\eta_2 \ge 1$ such that a feasible solution of SDP \eqref{eq:sdp_noise} can be constructed as $\eta_2 (\bar{\gamma}(k), \bar{Q}(k), \bar{P}(k), \bar{L}(k), \bar{Q}(k)\bar{P}(k)^{-1}\bar{Q}(k)^\prime)$, for $\bar{d} < \delta_{d,2}$ with $\delta_{d,2}$ defined in Lemma \ref{lem:relationship}. 
Here, $(\bar{\gamma}(k), \bar{Q}(k), \bar{P}(k), \bar{L}(k))$ is any optimal solution of SDP \eqref{eq:sdp_ideal}. Let  $(\gamma^*(k), Q^*(k), P^*(k), L^*(k),V^*(k))$ denote any optimal solution of SDP \eqref{eq:sdp_noise}. Then, it follows that
\begin{equation*}
{\rm tr}(P^*(k)) + {\rm tr}(L^*(k)) + \alpha {\rm tr}(V^*(k)) \le \eta_2 \bar{\gamma}(k) 
\end{equation*}
where $\bar{\gamma}(k) = \alpha {\rm tr}(\bar{Q}(k)\bar{P}(k)^{-1}\bar{Q}(k)^\prime) + {\rm tr}(\bar{P}(k)) + {\rm tr}(\bar{L}(k))$.
From the second constraint of SDP \eqref{eq:sdp_noise}, it can be observed that $P^*(k) \succeq I$. Hence, ${\rm tr}(L^*(k)) = {\rm tr}(K^*(k)P^*(k)K^*(k)^\prime) \ge {\rm tr}(K^*(k)K^*(k)^\prime) \ge \Vert K^*(k)\Vert^2$. Thus, it follows that
\begin{equation}\label{eq:K_bound1}
\Vert K^*(k)\Vert \le \sqrt{\eta_2 \bar{\gamma}(k) - n_x}.
\end{equation}

Here, since $k$ can be infinitely large, the inequality \eqref{eq:K_bound1} can result in an infinite number of upper bounds on $\Vert K^*(k)\Vert$. In the following, we demonstrate that by taking the modes of subsystems into account, there exist a finite number of $\bar{\gamma}(k)$.
As a consequence, $\Vert K^*(k)\Vert$ can be upper bounded by using the maximum one.

Let us consider any $j \in \mathbb{N}$ and $k = k^j = k_{s_j} + T$. 
The samples in matrices $X_{k - 1}$ and $X_k$ are generated from the same subsystem denoted by $i$, where $\sigma(k_{s_j}) = i \in\mathcal{M}$.
We denote the unique LQR control gain of system $(A_i, B_i)$ by $\bar{K}_i$, the controllability Gramian matrix by $\bar{P}_i$, and  $\bar{\gamma}_i := {\rm tr}(\bar{P}_i) + {\rm tr}(\bar{K}_i\bar{P}_i\bar{K}_i')$ the associated cost.
According to \cite[Lemma 4]{Rotulo2021Online}, the SDP \eqref{eq:sdp_ideal} is feasible and the matrix $\bar{K}(k) = U_{k - 1}\bar{Q}(k) \bar{P}(k)^{-1}$, where $(\bar{Q}(k), \bar{P}(k))$ is any optimal solution of SDP \eqref{eq:sdp_ideal}, is such that $\bar{K}(k) = \bar{K}_i$, $\bar{P}(k) = \bar{P}_i$, and $\bar{\gamma}(k) = \bar{\gamma}_i$. 
Moreover, observing from \eqref{eq:kstrategy}, \eqref{eq:P_strategy}, and Lemma \ref{lem:relationship} that $K(k) = K(k - 1)$ holds for all $k \in [k_{s_j} + T+ 1, k_{s_{j+1}}]$, we obtain from \eqref{eq:K_bound1} that for $k \in \cup_{j \in \mathbb{N}}[k_{s_j} + T, k_{s_{j+1}}]$, 
\begin{equation}\label{eq:K_bound2}
\Vert K(k)\Vert \le \max_{i \in \mathcal{M}} \sqrt{\eta_2 \bar{\gamma}_{i} - n_x}.
\end{equation}

Lemma \ref{lem:relationship} confirms that $k_j = k_{s_j}$. According to \eqref{eq:kstrategy}, for any $j \in \mathbb{N}$ and $k \in [k_{s_j} + N, k_{s_j} + T - 1]$, it has been shown in \cite[Lemma 5]{Rotulo2021Online} that if there are at least $N$ data generated from subsystem $\sigma(k_{s_j} - 1) = z$, then $(\bar{\gamma}_{z}, \bar{Q}_z, \bar{P}_{z}, \bar{K}_{z}\bar{P}_{z}\bar{K}_{z}')$  is a feasible solution of SDP \eqref{eq:sdp_ideal} with 
\[
\bar{Q}_z = W_{k-1}^\dag\left[\begin{matrix}\bar{K}_{z}\\I \end{matrix}\right] \bar{P}_z.
\]
Alternatively, if there are at least $N$ samples from subsystem $\sigma(k_{s_j}) = i$, then 		$(\bar{\gamma}_{i}, \bar{Q}_{i}, \bar{P}_{i}, \bar{K}_{i}\bar{P}_{i}\bar{K}_{i}')$ is a feasible solution of SDP \eqref{eq:sdp_ideal}.
Based on Lemma \ref{lem:feasible_2}, it can be concluded that inequality \eqref{eq:K_bound2} is valid for all $k$ within the intervals $\cup_{j \in \mathbb{N}}[k_{s_j} + N, k_{s_j} + T - 1]$.

Therefore, for all $k$ outside the range of $[k_j, k_j + N - 1]$, it can be established that  $\|K(k)\|$ is less than or equal to $\delta_K$. Here, $\delta_K$ can be defined as $\delta_K \triangleq \max_{i \in \mathcal{M}} \sqrt{\eta_2 \bar{\gamma}_{i} - n_x}$. This inequality completes the proof.
\end{pf}

The following stability result for system \eqref{eq:sys} is established.

\begin{theorem}
Consider the switched system \eqref{eq:sys} with controller \eqref{eq:uk}--\eqref{eq:P_strategy}. Let Assumptions \ref{as:sys}--\ref{as:dwell} hold. For all $k \in \mathbb{N}$, if $\bar{d} < \delta_{d,2}$, and $\lambda_0 \in [\check{\lambda}_0, \hat{\lambda}_0)$ with $\delta_{d,2}$, $\check{\lambda}_0$, and $\hat{\lambda}_0$ as in Lemma \ref{lem:relationship}, then there exist constants $\bar{\delta}_{\epsilon} >\underline{\delta}_{\epsilon} >0$, $\bar{\delta}_x>\underline{\delta}_x >0$, and $\bar{\tau} >0$ such that, for all $\delta_{\epsilon} \in (\underline{\delta}_{\epsilon}, \bar{\delta}_{\epsilon})$, $\delta_{x} \in (\underline{\delta}_x, \bar{\delta}_x)$, and $\tau \ge \bar{\tau}$, the closed-loop system achieves ISpS.
\end{theorem}

\begin{pf}
From \eqref{eq:uk}--\eqref{eq:P_strategy}, we observe that the controller gain matrix $K(k)$ stops updating when function $\mathcal{V}(x(k))$ is small, i.e., $ \mathcal{V}(x(k)) \le {\delta}_{V}$. 
Since $\Vert P(k)\Vert \le \max_{i \in \mathcal{M}} \eta_2 \bar{\gamma}_i$ as shown in the proof of Lemma \ref{lem:relationship}, this condition is equivalent to $\Vert x(k)\Vert \le \delta_{x}$ for some small $\delta_x >0$.
Consider the time instant $k_{in}$ when the state enters the range $\mathbb{B}_{\delta_{x}}$ for the first time, where $\mathbb{B}_{\delta_{x}}$ denotes a ball centered at the origin with radius $\delta_x$. Formally, $k_{in} := \min\{k\in \mathbb{N}\,|\, x(k) \in \mathbb{B}_{\delta_{x}}\}$.

The proof is divided into two steps with respect to $k_{in}$. First, we demonstrate that the state converges for all $k \in [0,k_{in})$. Then, we show that for all $k \ge k_{in}$, the state either converges to zero, or escapes from $\mathbb{B}_{\delta_{x}}$ and returns within finite time.

\emph{Step 1: State convergence before entering $\mathbb{B}_{\delta_{x}}$.}


Assuming that $\forall k \le T$, $\mathcal{V}(x(k)) > \delta_V$, and $x(k) \notin \mathbb{B}_{\delta_{x}}$, we consider an arbitrary time interval $[k_{s_j}, k_{s_j+1} - 1] \subseteq [0,k_{in})$. Let $i = \sigma(k_{s_j})$ denote the activated subsystem for $k \in [k_{s_j}, k_{s_j+1} - 1]$. 
We now show that the system achieves stability for all $k \in [k_{s_j} + T, k_{s_{j+1}} - 1]$.
According to Lemmas \ref{lem:feasible_1} and \ref{lem:relationship}, it holds for $\lambda_0 \in [\check{\lambda}_0, \hat{\lambda}_0)$ as in Lemma \ref{lem:relationship}, that $\mathcal{W}_i(x(k + 1)) \le \lambda_0\mathcal{W}_i(x(k))$, and recursively
\begin{equation*}
{\mathcal{W}_i(x(k)) \le \lambda_0^{k - k_{s_j} - T}\mathcal{W}_i(x(k_{s_j} + T))}
\end{equation*}
then
\begin{equation}\label{eq:x:con}
\Vert x(k)\Vert \le \phi_0\tilde{\lambda}_0^{k - k_{s_j} - T}\Vert x(k_{s_j} + T)\Vert
\end{equation}
where $\phi_0 := [(\max_{i \in \mathcal{M}} \eta_2 \bar{\gamma}_i)/{\underline{\lambda}_P}]^{1/2}$
and $\tilde{\lambda}_0 := \sqrt{\lambda_0}$.

Based on \eqref{eq:uk} and Theorem \ref{thm:k_bound}, the growth of states between two consecutive times can be upper bounded as follows
\begin{align}
\Vert x(k + 1)\Vert &\le \max\!\Big\{\Vert A_{\sigma(k)} + B_{\sigma(k)} K(k)\Vert \Vert x(k)\Vert + \bar{d},\nonumber\\
&\quad  \Vert A_{\sigma(k)}\Vert \Vert x(k)\Vert + \Vert B_{\sigma(k)}\Vert  \delta_{\epsilon}+ \bar{d}\Big\}\\
&\le C \Vert x(k)\Vert + \bar{d}\label{eq:C}
\end{align}		
where $C := \max\{C_0, C_1, 1 \}$ with $C_0 := \max_{i \in \mathcal{M}}(\Vert A_i\Vert + \Vert B_i\Vert\delta_{K})$ and $C_1 :=\max_{i \in \mathcal{M}} (\Vert A_i\Vert + \Vert B_i\Vert\delta_{\epsilon}/\delta_{x})$.

Noting inequality \eqref{eq:C} holds for all $k \in [k_{s_j} + 1, k_{s_j} + T]$. 
Combining this fact with \eqref{eq:x:con}, we obtain the following inequality for all $t \in [1, T]$
\begin{align}
\Vert x(k_{s_j} + t) \Vert &\le C^t \phi_0^s \tilde{\lambda}_0^{k_{s_j} - k_0}\Big(\frac{C}{\lambda_0}\Big)^{sT} \Vert x(k_0) \Vert + \bar{d} \sum_{i = 0}^{t - 1}C^i \nonumber\\
&~~~+\frac{ C^{t + T}}{C - 1}\bar{d} \sum_{i = 1}^{j} \phi_0^{i} \tilde{\lambda}_0^{k_{s_j} - k_{s_j - i}}\Big(\frac{C}{\tilde{\lambda}_0}\Big)^{iT} \label{eq:V1}.
\end{align}
For $t \in [T + 1, k_{s_{j+1}} - k_{s_j}]$, we have that
\begin{align}
&\Vert x(k_{s_j} + t) \Vert \le \tilde{\lambda}_0^t\phi_0^{s + 1} \tilde{\lambda}_0^{k_{s_j} - k_0}\Big(\frac{C}{\tilde{\lambda}_0} \Big)^{(s + 1)T}\Vert x(k_0) \Vert \nonumber\\
&\qquad \qquad ~ + \frac{\lambda_0^t C^T}{C - 1}\bar{d} \sum_{i = 0}^{j}\phi_0^{i + 1}\tilde{\lambda}_0^{k_{s_j} - k_{s_j - i}}\Big(\frac{C}{\tilde{\lambda}_0}\Big)^{(i + 1)T}\label{eq:V2}.
\end{align}

Choose  constant $\mu$ such that $0 < \tilde{\lambda}_0< \mu <1$. 
Let $\tau > \bar{\tau}$ with
\begin{equation}
\bar{\tau} :=\frac{\ln({C}/{\tilde{\lambda}_0})}{\ln(\phi_0({\mu}/{\tilde{\lambda}_0})^T)}.
\end{equation}
Hence, for all $t \in [1, T]$, it follows that
\begin{align*}
&C^t \phi_0^s \tilde{\lambda}_0^{k_{s_j} - k_{s_0}}\Big(\!\frac{C}{\lambda_0}\Big)^{jT}\!\!\\
=&~ \mu^{t \!+\! k_{s_j} \!- k_0} \!\Big(\frac{C}{\mu} \Big)^t\!\Big(\frac{\tilde{\lambda}_0}{\mu} \Big)^{k_{s_j} - k_{s_0}}\!\phi_0^j\Big(\frac{C}{\tilde{\lambda}_0} \Big)^{jT}\\
\!\! \le&~ \mu^{t + k_{s_j} - k_{s_0}} \Big(\frac{C}{\mu} \Big)^T\Big(\frac{\tilde{\lambda}_0}{\mu} \Big)^{j\tau}\phi_0^j\Big(\frac{C}{\tilde{\lambda}_0} \Big)^{jT}\\
 \le&~ \phi_0 \Big(\frac{C}{\mu} \Big)^T\mu^{t + k_{s_j} - k_{s_0}}.
\end{align*}
Similarly, for all $t \in [T + 1, k_{s_{j+1}} - k_{s_j}]$, it holds that
\begin{align*}
&\tilde{\lambda}_0^t\phi_0^{j + 1} \tilde{\lambda}_0^{k_{s_j}\!\!- k_{s_0}}\!\Big(\frac{C}{\tilde{\lambda}_0} \Big)^{\!(j + 1)T} \!\!\!\!\\
\le&~ \mu^{k_{s_j} \!+ t - k_{s_0}}\Big(\frac{\tilde{\lambda}_0}{\mu} \Big)^{\!j\tau}\!\!\!\phi_0^{j + 1}\Big(\frac{C}{\tilde{\lambda}_0} \Big)^{\!(j + 1)T}\\
 \le&~ \phi_0\Big(\frac{C}{\mu} \Big)^T\mu^{k_{s_j} + t - k_{s_0}}.
\end{align*}
Therefore, it can be deduced from \eqref{eq:V1} and \eqref{eq:V2} that 
\begin{align*}
\Vert x(k_{s_j} + t))\Vert \le \phi_0\Big(\frac{C}{\mu} \Big)^T\mu^{t + k_{s_j} - k_{s_0}} \Vert x(k_{s_0})\Vert + \varphi(\bar{d})
\end{align*}
where $\varphi(\cdot) \in \mathcal{K}$,
which implies that for sufficiently small $\bar{d}$, the state converges.
Hence, for $\delta_{x} > 0$, there exists a time $0 < k_{in} < \infty$ such that $x(k_{in}) \in \mathbb{B}_{\delta_{x}}$.

\emph{Step 2: State convergence after first entering $\mathbb{B}_{\delta_{x}}$}.

Suppose that for some switching time $k_{s'_j} \in \mathbb{N}$ and $k_{in} \in [k_{s_j^\prime} + T, k_{s_j^\prime + 1}]$ for some $k_{s_j^\prime}$, the mode of the switched system remains unchanged, i.e., $j = \sigma(k) = \sigma(k_{in})$ for all $k \ge k_{in}$. Noticing that the disturbance satisfies $\bar{d} < \delta_{d,2}$, the Lyapunov function decreases with at least rate $\lambda_0$. Therefore, $\mathcal{W}_j(x(k)) \le \lambda_0 \mathcal{W}_i(x(k - 1))$  for $k \ge k_{in}$. Recursively, the state satisfies $\Vert x(k)\Vert \le \phi_0\tilde{\lambda}_0^{(k - k_{in})/2}\delta_{x}$, indicating that the state converges to the origin as $k \rightarrow \infty$.

Next, we show that if the state escapes from the range $\mathbb{B}_{\delta_{x}}$, it will come back within finite times. For $k_{in} \in [k_{s_j^\prime} + 1, k_{s_j^\prime} + T - 1]$, and $\sigma(k_{s_j^\prime + 1}) \ne \sigma(k_{s_j^\prime})$, from \eqref{eq:C}, the state satisfies $\Vert x(k_{in} + 1)\Vert = C\Vert x(k_{in})\Vert + \bar{d}$ with $C \ge 1$. This implies that the state may diverge. Therefore, there exists a time $k_{out} \in (k_{in}, \infty)$ such that $x(k_{out}) \notin \mathbb{B}_{{\delta}_{x}}$, and the state feedback controller gain $K(k_{out})$ continues to be updated. It follows from {Step 1} that the state gets back to this range within finite times.

In conclusion, the state satisfies
\begin{equation*}
|x(k_{s_j} + t) | \le \phi_0\Big(\frac{C}{\mu} \Big)^T\mu^{t + k_{s_j} - k_{s_0}} \Vert x(k_{s_0})\Vert + \varphi(\bar{d}) + \delta_{x},
\end{equation*}
which completes the proof according to Definition \ref{def:isps}.
\end{pf}

It is worth remarking that the disturbance-free data-driven controller proposed in \cite{Rotulo2021Online} cannot be directly extended to handle cases with disturbances, as we show below.
\begin{remark}[Relative to \cite{Rotulo2021Online}]
	\label{rem:relative}
In the absence of disturbances, the controller in \cite{Rotulo2021Online} utilizes a perturbed feedback control law given by $	u(k) = K(k)x(k) + \bar{\epsilon}(k)\Vert x(k)\Vert$, where $K(k)x(k)$ is added to an auxiliary signal $\bar{\epsilon}(k)\Vert x(k)\Vert$ with $\bar{\epsilon}(k)$ selected within the ball $\mathbb{B}_{\delta_{\bar{\epsilon}}} := \{\bar{\epsilon}\in \mathbb{R}^{n_u}\,|\,\Vert \epsilon\Vert \le \delta_{\bar{\epsilon}} \}$. It has been proven in \cite[Lemma 3]{depersis2020Low} that the rank condition \eqref{eq:rankWk} can be guaranteed at every time $k$ by the auxiliary signal $\bar{\epsilon}(k)$, ensuring the feasibility of SDP \eqref{eq:sdp_ideal} for all $k \in \mathbb{N}$. Building on this fact, the disturbance-free system can be stabilized with the perturbed controller by taking $K(k) = U_{k - 1}\bar{Q}(k)\bar{P}(k)^{-1}$, where $(\bar{P}(k), \bar{Q}(k))$ are optimal solutions of SDP \eqref{eq:sdp_ideal} at each time $k$. 		Intuitively, one might be tempted to directly apply this controller when disturbances are present. In other words, if $\bar{\epsilon}(k)$ can guarantee the rank condition \eqref{eq:rankWk} for any $k$ with $d(k)\neq 0$, then we can adopt the controller and update $K(k)$ by solving SDP \eqref{eq:sdp_noise}. However, as we illustrate through an example, the condition \eqref{eq:rankWk} can be violated when nonzero disturbance $d(k)$ is involved.

Consider the case where $(A_{\sigma(k)}, B_{\sigma(k)}) = (1,1/2)$ for all $k \in [0,4]$ in \eqref{eq:sys} and $x(0) = 1$. Under the assumption that $T = 2((n_x + 1)n_u + n_x)-1 = 5$ (cf. Assumption \ref{as:T}), let us take $\bar{\epsilon}(k) = -1/2$ and $K(k) = -1$ for all $k \in [0,4]$ for simplicity.
In the absence of disturbances, i.e., $d(0) = d(1) = d(2) = d(3) = 0$, we can obtain the input-state sequence $({x}(0),u(0)) = (1,-3/2)$, $({x}(1),u(1)) = (1/4,-3/4)$, $({x}(2),u(2)) = (-1/8,1/16)$, $({x}(3),u(3)) = (-3/32,3/64)$, $({x}(4),u(4)) = (-9/128, 9/256)$. As a result, the matrix $W_5$ defined by
\begin{equation*}
W_{5} = \left[
\begin{matrix}
	U_5\\
	X_5
\end{matrix}
\right] = 
\left[
\begin{matrix}
	-{3}/{2}\! &-{3}/{4}\! & {1}/{16}\!& -{3}/{64}\!& -{9}/{256}\\
	1 \!&{1}/{4}\!&-{1}/{8}\!&-{3}/{32}\!&-{9}/{128}
\end{matrix}
\right]
\end{equation*}
has full row rank, and hence condition \eqref{eq:rankWk} is satisfied.

However, when the disturbance sequence is nonzero, with e.g., $d(0) = 3/4$, $d(1) = (1/8)$, $d(2) = 5/48$, and $d(3) = 1/12$, while retaining the same initial condition $x(0) = 1$ and variables $\epsilon(k) = -1/2$, $K(k) = -1$ for all $k \in [0,4]$, the input-state sequence becomes $({x}(0),u(0)) = (1,-3/2)$, $({x}(1),u(1)) = (1/2,-3/4)$, $({x}(2),u(2)) = (1/4,-3/8)$, $({x}(3),u(3)) = (1/6,-1/4)$, and $({x}(4),u(4)) = (1/8, -3/16)$. In this case, the matrix $W_5$ given by
\begin{equation*}
W_{5} = \left[
\begin{matrix}
	U_5\\
	X_5
\end{matrix}
\right] = 
\left[
\begin{matrix}
	-{3}/{2}\! &-{3}/{4}\! & -{3}/{8}\!& -{1}/{4}\!& -{3}/{16}\\
	1\!&{1}/{2}\!&{1}/{4}\!&{1}/{6}\!&{1}/{8}
\end{matrix}
\right]
\end{equation*}
is such that ${\rm rank}(W_5) = 1$, violating condition \eqref{eq:rankWk}. Therefore, SDP \eqref{eq:sdp_noise} is not feasible.
\end{remark}

\section{Numerical Examples}
\label{sec:example}

In this section, we examine two numerical examples to validate the effectiveness of the proposed controller, both of which have been previously considered in \cite{Rotulo2021Online} under the disturbance-free case.

\subsection{Flight control system}

The first example considers the stabilization problem of the linearized longitudinal dynamics of an F-18 aircraft operating at Mach $0.3$ and altitude $26$ kft, and Mach $0.7$ and altitude $14$ kft, respectively. This problem can be described using two linear subsystems  $(A_1, B_1)$ and $(A_2, B_2)$. Both subsystems are controllable and have the following matrices:
\begin{align*}
A_1 &= \left[
\begin{matrix}
0.977 & 0.097\\
0.002 & 0.981
\end{matrix}
\right],~~~ B_1 = \left[
\begin{matrix}
-0.013 & -0.004\\
-0.171 & -0.051
\end{matrix}
\right],\\
A_2 &= \left[
\begin{matrix}
0.852 & 0.088\\
-0.753 & 0.87
\end{matrix}
\right],~ B_2 = \left[
\begin{matrix}
-0.106 & -0.021\\
-1.8143 & -0.358
\end{matrix}
\right].
\end{align*}

We set $T = 15$, $\bar{d} = 0.03$, $\delta_{V} = 0.05$, $\alpha = 1$, and $\lambda_0 = 0.945$. We generate an arbitrary switching signal $\sigma$ with dwell time $\tau \geq 15$. We first collect an offline input-state trajectory by applying a sequence of inputs $u$ uniformly distributed in $[-0.3, 0.3]$ to the subsystem $(A_1, B_1)$ only. Additionally, we prepare an $\mu$-persistently exciting input sequence $\{u(k)\}_{k = 0}^{N - 1}$ of order $3$ with $\mu = 0.01$. We then run the system online using the proposed control strategy \eqref{eq:uk}--\eqref{eq:P_strategy}.

\begin{figure}
\centering
\includegraphics[width=9cm]{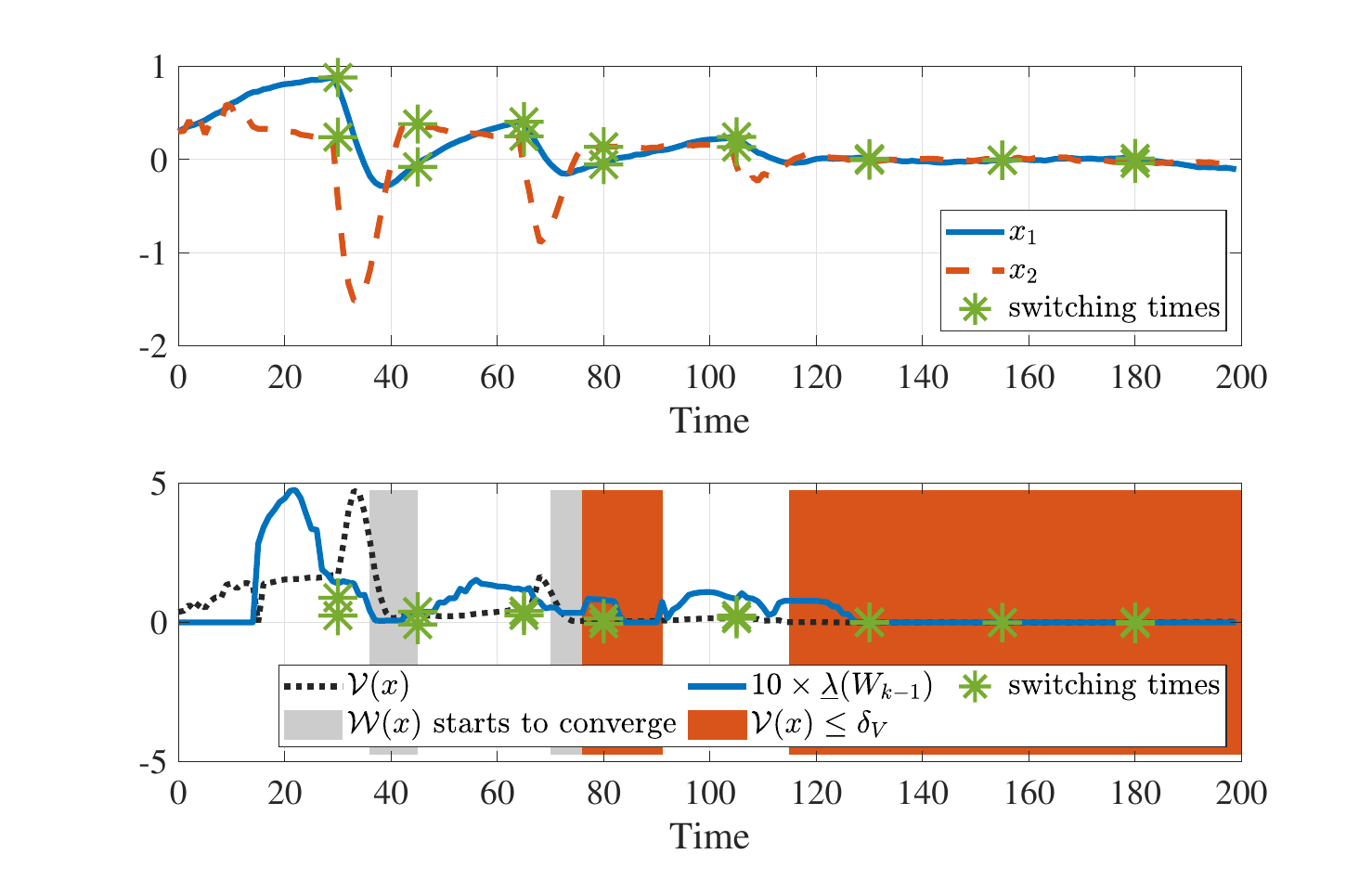}\\
\caption{State-input trajectory of flight control systems: disturbance-free data with the controller in \cite{Rotulo2021Online} (top two panels) and noisy data with the proposed controller \eqref{eq:uk} (bottom two panels).}\label{fig:xu}
\centering
\end{figure}

The top panel of Fig. \ref{fig:xu} depicts the convergence of the state trajectory over a simulation horizon of $200$ time instants. The offline data-collection phase is shown in the interval $t \in [0,15]$. The bottom panel of Fig. \ref{fig:xu} shows the evolution of the Lyapunov function and the smallest singular value of the matrix $W_{k - 1}$. The gray shades represent the phase when the auxiliary function $\mathcal{V}(x(k))$ converges with rate $\lambda_0$, indicating the convergence of the Lyapunov function $\mathcal{W}(x(k))$. The orange shades indicate the phase when $\mathcal{V}(x(k)) \leq \delta_V$. We observe that at $k = 76$, function $\mathcal{V}(x(k))$ first enters the range $\mathbb{B}_{\delta_V}$ (state $x(k)$ entering the range $\mathbb{B}_{\delta_x}$). However, the system switches at $k = 80$, causing the state to escape from this range. Nonetheless, the proposed control strategy \eqref{eq:uk}--\eqref{eq:P_strategy} eventually causes the state to converge to a small range.

\begin{figure}
\centering
\includegraphics[width=9cm]{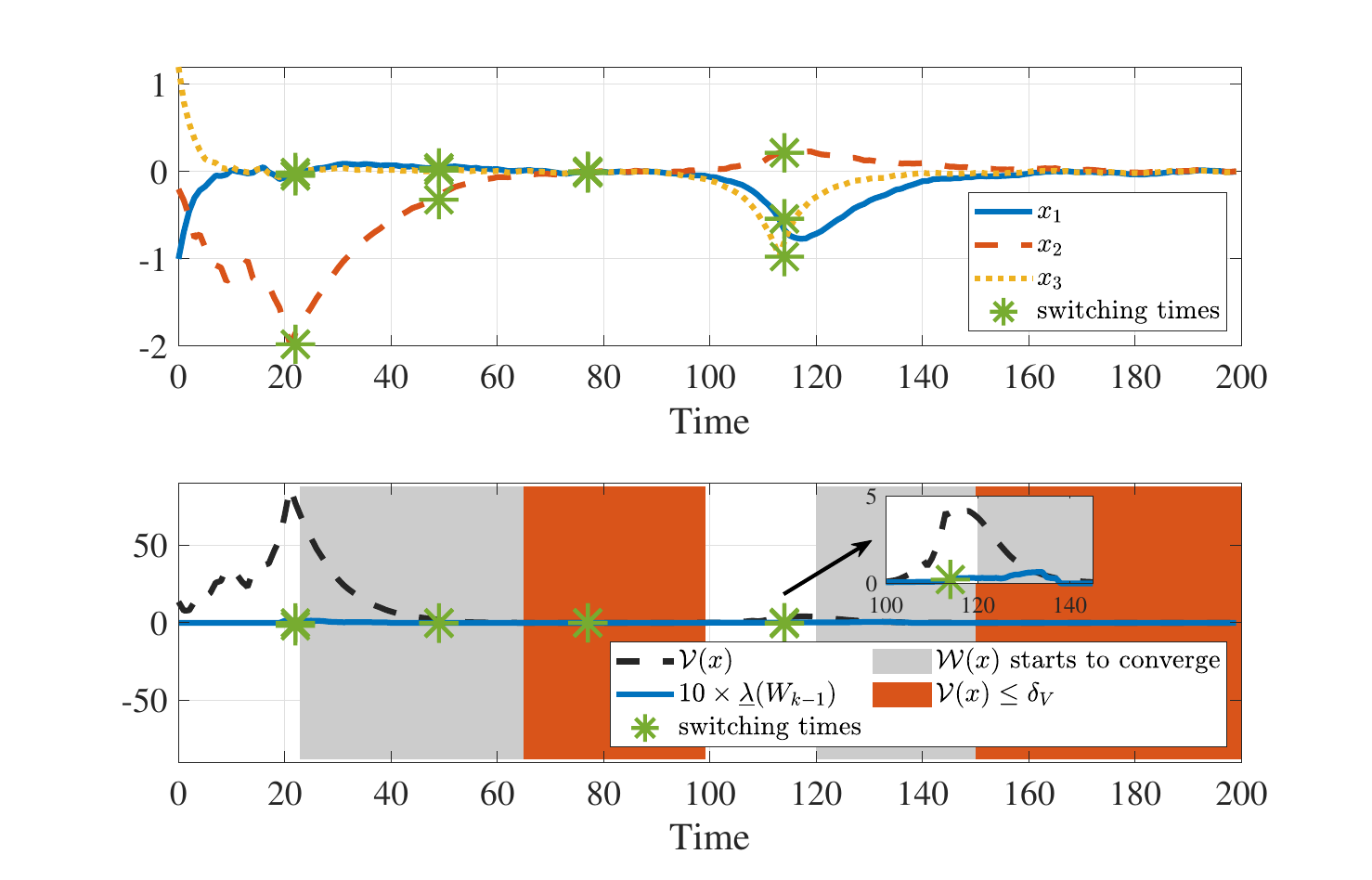}\\
\caption{State-input trajectory of aircraft engine systems: disturbance-free data with the controller in \cite{Rotulo2021Online} (top two panels) and noisy data with the proposed controller \eqref{eq:uk} (bottom two panels).}\label{fig:xu3}
\centering
\end{figure}
\subsection{Aircraft engine systems}

In the second example, we consider our approach for stabilizing fault tolerant systems as switched systems. Specifically, we apply our proposed online controller to an F-$404$ aircraft engine system subject to system and actuator faults, originally considered in \cite{Liu2017sliding}. The system is composed of three states, namely, the sideslip angle, the roll rate, and the yaw rate, with two control inputs representing the engine thrust and the flight path angle. We consider a discretized linearized version of this system with a sampling period of $0.1$s, and the system matrices are given by 	\begin{align*}
A = \left[
\begin{matrix}
0.867 & 0& 0.202\\
0.015 &0.961 &-0.032\\
0.026& 0& 0.803\\
\end{matrix}
\right],~B = \left[
\begin{matrix}
0.011& 0\\
0.014 &-0.039\\
0.009& 0\\
\end{matrix}
\right].
\end{align*} 

Similarly to the previous section, we collect an input-state trajectory of length $T=21$ by simulating the system offline using a persistently exciting input sequence uniformly generated in $[-3.5, 3.5]$. Fig. \ref{fig:xu3} shows the offline phase represented by the interval $t\in [0,21]$. We then run the system online, where external disturbances and unknown faults such as wind gusts or structural vibrations can undermine system stability, characterized by changes in the system matrices $A$ and $B$. Specifically, system faults are captured by the changes in the system matrices as $\tilde{A} = A + \beta(k)D$ with
\begin{align*}
D = \left[
\begin{matrix}
0.075 & 0 & 0\\
0.5 & 1 & 0\\
0 &0  &-0.75
\end{matrix}
\right],~\beta(k) = \begin{cases}
0.1& k \in [0,27]\\
0.05 & k \in [27,52]\\
-0.5 & k\in [52,95]\\
0 & {\rm else}
\end{cases}
\end{align*}  while failures of the engine generating thrust and the motor moving the path angle are modeled by the changes in the input matrix $B$ as $\tilde{B} = B\alpha(k)$ with
\begin{align*}
\alpha(k) = 
\left[
\begin{matrix}
1&0\\
0&0
\end{matrix}
\right],\forall k \in [27,52],~~~ \alpha(k) =
\left[
\begin{matrix}
0&0\\
0&1
\end{matrix}
\right],\forall k \ge 52.
\end{align*}

During online operation, we track the state trajectory (top panel), evolution of the function $\mathcal{V}(x(k))$, and the smallest singular value of $W_{k-1}$ (bottom two panels) as shown in Fig. \ref{fig:xu3}. We set $\bar{d} = 0.015$, $\delta_u = 0.01$, $\delta_{x} = 10^{-4}$, and $\alpha = 0.2$.


\section{Acknowledgments} 
The authors would like to thank Prof. Claudio De Persis for his kind help and valuable suggestions on Remark \ref{rem:relative} and the paper's structure.

\section{Conclusions}\label{sec:conclusion}

In this paper, we presented a data-driven switched controller for stabilizing unknown linear switched systems, utilizing noisy input-state data. Our approach employs an auxiliary function-based switching law, where the state feedback gain is updated by solving a robust data-based SDP online. We provided conditions such that the feasibility of the robust SDP is guaranteed, and established the ISpS under the assumptions that the disturbance is bounded and the system switches slowly enough. Two numerical examples were employed to demonstrate the practical benefits of our proposed controller.

\renewcommand{\thesection}{Appendix A}

\subsection{Proof of Theorem \ref{lem:wpe}}
\label{sec:app:lem:wpe}

\begin{pf}
\label{pf:lem:wpe}
Partition matrix $W_{-1}$ in \eqref{eq:X0:W} into
$W_{-1} := W_{-1,\bar{x}} + W_{-1,d}$ with
\begin{equation}\label{eq:W0:par}
	W_{-1,\bar{x}} := \left[\begin{matrix}
		U_{-1}\\
		\bar{X}_{-1}
	\end{matrix}\right],\quad W_{-1,d} := \left[\begin{matrix}
		0\\
		\tilde{D}_{-1}
	\end{matrix}\right]
\end{equation}
where $\bar{X}_{-1} = [\bar{x}( - T)~ \cdots~\bar{x}( - 1)]$ with $\bar{x}( - T) = x( - T)$ and subsequent ones computed by the disturbance-free recursion
$
\bar{x}(i + 1) = A\bar{x}(i) + Bu(i)$ for all $ i  \in[- T, -2]
$,
and, where $\tilde{D}_{-1,p} := \sum_{i = 0}^{p - 1}\Pi_{j = 1}^{i}A d( T + p - i - 1)$ is the $p$-th column of matrix $\tilde{D}_{-1}$.
Since $\Vert D_{-1} \Vert \le \delta$, it can be deduced from \eqref{eq:sys_lti_noise} that $\Vert \tilde{D}_{-1}\Vert \le \sqrt{n_x} \Vert \Omega_0\Vert\Vert D_{-1}\Vert\le \sqrt{n_x} \Vert \Omega_0\Vert \delta$ with
\begin{align*}
	\Omega_0 := \left[
	\begin{matrix}
		0 & 0 & 0 & \cdots & 0\\
		I & 0 & 0 & \cdots & 0\\
		A & I & 0 & \cdots & 0\\
		\vdots & \vdots &\vdots & \ddots & \vdots\\
		A^{T - 2} & A^{T - 3} & A^{T - 4} & \cdots & 0
	\end{matrix}
	\right].
\end{align*}
On the other hand, it follows from \cite[Theorem 3.1]{coulson2022robust} that for $\bar{w}$-persistently exciting input sequence $u(-T),\cdots,u(-1)$, the smallest singular value of matrix $W_{-1, \bar{x}}$ satisfies $\underline{\lambda}_{W_{-1, \bar{x}}} \ge \bar{w}\rho/\sqrt{n_x + 1}$ where $\rho >0$ is an internal parameter of system $(A,B)$; see \cite[Lemma 3.1]{coulson2022robust} for its detailed definition.
If constant $\bar{w}$ satisfies
\begin{equation}
	\bar{w} > \sqrt{n_x(n_x + 1)}\Vert \Omega_0 \Vert\delta/\rho
\end{equation}
then it holds that $\underline{\lambda}_{W_{-1,\bar{x}}} >\sqrt{n_x} \Vert \Omega_0\Vert \delta \ge \Vert \tilde{D}_{-1}\Vert = \Vert W_{-1,d} \Vert$.
Recall the fact that for any matrices $Y$, $Z$ of the same size, inequality $\underline{\lambda}_{Y+ Z}\ge \underline{\lambda}_{Y} - \Vert Z\Vert$ holds true.
Hence, one has from \eqref{eq:W0:par} that $\underline{\lambda}_{W_{-1}} \ge \underline{\lambda}_{W_{-1,\bar{x}}} - \Vert W_{-1,d} \Vert >0$ indicating that condition \eqref{eq:rankW0} holds.
\end{pf}

\subsection{Proof of Theorem \ref{lem:feasible_1}}
\label{sec:lem:feasible_1}
\begin{pf}
\label{pf:lem:feasible_1}
	Partition matrix $W_{k - 1}$  following the same step as in the proof of Lemma \ref{lem:wpe}, i.e., $W_{k - 1} := W_{k - 1,\bar{x}} + W_{k - 1,d}$ with
	\begin{equation*}
		W_{k - 1,\bar{x}} := \left[\begin{matrix}
			U_{k - 1}\\
			\bar{X}_{k - 1}
		\end{matrix}\right],\quad W_{k - 1,d} := \left[\begin{matrix}
			0\\
			\tilde{D}_{k - 1}
		\end{matrix}\right]
	\end{equation*}
	where $\bar{X}_{k - 1} = [\bar{x}(k - T)~ \cdots~\bar{x}(k - 1)]$ with $\bar{x}(k - T) = x(k - T)$ and subsequent ones computed by the disturbance-free recursion
	$
	\bar{x}(i + 1) = A_{\sigma(i)}\bar{x}(i) + B_{\sigma(i)}u(i)$ for all $ i  \in[k - T, k - 2]$,
	and,  where $\tilde{D}_{k - 1,p} := \sum_{i = 0}^{p - 1}\Pi_{j = 1}^{i}A_{\sigma(k - T + j)}d(k - T + p - i - 1)$ is the $p$-th column of matrix $\tilde{D}_{k - 1}$.

	Since the input sequence ${u(k - T), \cdots, u(k - 1)}$ is $\bar{w}$-persistently exciting of order $n_x + 1$, it follows from the proof of Lemma \ref{lem:wpe} that $\underline{\lambda}_{W_{k - 1, \bar{x}}} \ge \hat{w} := \bar{w} \rho/\sqrt{n_x + 1}$ for some constant $\rho >0$.
	In addition, note that $\Vert W_{k - 1,d}\Vert \le \sqrt{n_x}\Vert \Omega_{k - 1}\Vert \Vert D_{k - 1}\Vert$ where $\Omega_{k -1}$ is defined in \eqref{eq:omegak-1}, presented at the top of the next page.
	Since the system switches in a finite set, there exists a constant $\bar{\Omega}$ such that $\Vert\Omega_{k} \Vert \le \bar{\Omega}$ for all $k \in \mathbb{N}$.
	Moreover, it can be deduced from $\Vert D_{k - 1}\Vert \le T\bar{d}^2$ that $\Vert W_{k - 1,d}\Vert \le \sqrt{T n_x}\bar{\Omega}\bar{d}$ for all $k \in \mathbb{N}$.
	Therefore, if  
	\setcounter{equation}{26} 
	\begin{equation}\label{eq:uppbound_d1}
		\bar{d} < {\hat{w}}/(2\sqrt{Tn_x}\bar{\Omega})
	\end{equation}
	then $\underline{\lambda}_{W_{k -1, \bar{x}}} > \Vert W_{k -1, {d}}\Vert$ and $\underline{\lambda}_{W_{k -1}} \ge \underline{\lambda}_{W_{k -1, \bar{x}}} - \Vert W_{k -1, {d}}\Vert >0$ implying that condition \eqref{eq:rankWk} holds.
	Hence, SDPs \eqref{eq:sdp_noise} and \eqref{eq:sdp_ideal} are feasible, where the feasibility of SDP \eqref{eq:sdp_ideal} is guaranteed by \cite[Lemma 4]{Rotulo2021Online}.
	Let $(\bar{\gamma}(k), \bar{Q}(k), \bar{P}(k),\bar{L}(k))$ denote an optimal solution of SDP \eqref{eq:sdp_ideal}.
	It follows from \cite[Lemma 7]{depersis2020Low} that for some given constant $\eta_2 \ge 1$, if $-\bar{\Psi}(k) \le (1 - {1}/{\eta_2})I$ with $\bar{\Psi}(k) = D_{k - 1} \bar{M}(k)D_{k - 1}^\prime - X_k\bar{M}(k)D_{k - 1}^\prime - D_{k - 1}\bar{M}(k)X_k^\prime$ and $\bar{M}(k) = \bar{Q}(k)\bar{P}(k)^{-1}\bar{Q}(k)^\prime$, then SDP \eqref{eq:sdp_noise} is feasible, and any optimal solution $(\gamma^*(k), Q^*(k),P^*(k), L^*(k), V^*(k))$ constructs $K^*(k) = U_{k - 1}Q_{i}^*(k)P_{i}^*(k)^{-1}$ resulting to a Schur stable matrix $A_i+B_i K^*(k)$.
	
	\newcounter{TempEqCnt} 
	\setcounter{TempEqCnt}{\value{equation}} 
	\setcounter{equation}{25} 
	\begin{figure*}[!t]
		\normalsize
		\begin{equation}\label{eq:omegak-1}
			\Omega_{k - 1} = \left[
			\begin{matrix}
				0 & 0 & 0 & \cdots & 0\\
				I & 0 & 0 & \cdots & 0\\
				A_{\sigma(k - T + 1)} & I & 0 & \cdots & 0\\
				\vdots & \vdots & \vdots & \ddots & \vdots\\
				\Pi_{j = 1}^{T - 2}A_{\sigma(k - T + j)} &\Pi_{j = 0}^{T - 3}A_{\sigma(k - T + j)}& \Pi_{j = 0}^{T - 4}A_{\sigma(k - T + j)} & \cdots & 0
			\end{matrix}\right]
		\end{equation}
		\setcounter{equation}{\value{MYtempeqncnt}}
		\hrulefill
		\vspace*{4pt}
	\end{figure*}
	\setcounter{equation}{\value{TempEqCnt}} 
	
	In the following, a bound on the disturbance is derived ensuring that $-\bar{\Psi}(k) \le (1 - {1}/{\eta_2})I$ always holds true.
	Define $\bar{\Phi}(k) = [\bar{K}(k)^\prime~ I]^\prime$, where $\bar{K} = U_{k - 1}\bar{Q}(k)\bar{P}(k)^{-1}$.
	Combining $\bar{M}(k) = \bar{Q}(k)\bar{P}(k)^{-1}\bar{Q}(k)^\prime$ with $\bar{P}(k)^{-1} \succeq 0$, one gets that $\bar{M}(k) \succeq 0$, and hence a sufficient condition for $-\bar{\Psi}(k) \le (1 - {1}/{\eta_2})I$ is
	\begin{equation*}
		X_k\bar{M}(k)D_{k - 1}^\prime + D_{k - 1}\bar{M}(k)X_k^\prime \le (1 - {1}/{\eta_2})I.
	\end{equation*}
	Taking the $2$-norm for both sides of the above inequality, if 
	\begin{equation}\label{eq:sufficient_d1}
		2\Vert X_k\Vert \Vert \bar{M}(k)\Vert\Vert D_{k - 1}^\prime\Vert \le 1 - {1}/{\eta_2}
	\end{equation}
	then $-\bar{\Psi}(k) \le (1 - {1}/{\eta_2})I$.

	Combining $X_{k - 1}\bar{Q}(k) =\bar{P}(k)$ (the second constraint in \eqref{eq:sdp_ideal}) with $\bar{K}(k) = U_{k - 1} \bar{Q}(k)\bar{P}(k)^{-1}$, also considering that $W_{k - 1}$ has full row rank, matrix $\bar{Q}(k)$ can be expressed as $\bar{Q}(k) = W_{k - 1}^{\dag}\bar{\Phi}(k)\bar{P}(k)$.
	Hence, $\bar{M}(k) = W_{k - 1}^\dag\bar{\Phi}(k)\bar{P}(k)\bar{\Phi}^\prime(W_{k - 1}^\dag)^\prime$, and $\Vert \bar{M}(k)\Vert \le \Vert \bar{\Phi}(k)\Vert^2\Vert \bar{P}(k)\Vert\Vert W_{k - 1}^\dag\Vert^2$.
	Noticing from \cite[Theorem 4]{persis2020data} that $\bar{K}(k)$ is the unique solution of $(A_i + B_i\bar{K}(k))\bar{P}(k)(A_i + B_i \bar{K}(k))' - \bar{P}(k) + I = 0$,  there exists a constant $\phi$ such that $\Vert \bar{\Phi}(k)\Vert^2\Vert \bar{P}(k)\Vert \le \phi$ holds for all $i \in \mathcal{M}$.
	In addition, since $\underline{\lambda}_{W_{k - 1, \bar{x}}} = \Vert W_{k - 1,\bar{x}}^\dag \Vert^{-1}$ and $\Vert W_{k - 1}^\dag \Vert^{-1} \ge \Vert W_{k - 1,\bar{x}}^\dag \Vert^{-1} - \Vert W_{k - 1,d} \Vert$, 
	under condition \eqref{eq:uppbound_d1}, inequality $\Vert W_{k - 1}^\dag \Vert \le 2\Vert W_{k - 1, \bar{x}}^\dag\Vert$ holds.
	Furthermore, since columns in $X_{k}$ are generated by subsystem $(A_i, B_i)$, it follows that $X_k = [B_{i}~A_{i}]W_{k - 1} + D_{k - 1}$. 
	Based on the fact that $\Vert W_{k - 1}\Vert \le \Vert W_{k - 1,\bar{x}}\Vert + \sqrt{T n_x}\bar{\Omega}\bar{d}$, inequality 
	\eqref{eq:sufficient_d1} becomes
	\begin{align*}
		&8\sqrt{T}\bar{d}\phi \Vert W_{k - 1,\bar{x}}^\dag\Vert^2(\Vert \Xi_i\Vert(\Vert W_{k - 1,\bar{x}} \Vert + \sqrt{T n_x}\bar{\Omega}\bar{d}))\\
		& + 8\sqrt{T}\bar{d}\phi \Vert W_{k - 1,\bar{x}}^\dag\Vert^2\sqrt{T}\bar{d} \le 1 - {1}/{\eta_2}
	\end{align*}
	where $\Xi_i := [A_i~B_i]$.
	Since the switched system has finite modes, there exists a constant $\bar{\Xi}$ such that $\Vert \Xi_i\Vert \le \bar{\Xi}$ for all $i \in \mathcal{M}$.
	Considering \eqref{eq:uppbound_d1}, a sufficient condition for \eqref{eq:sufficient_d1} is 
	\begin{align}\label{eq:d_bound1}
		\bar{d} <\delta_{d,1}  := \hat{w} \min\bigg\{&\frac{1}{2\sqrt{T n_x}\bar{\Omega}},\frac{1 - 1/\eta_2}{24\sqrt{T}\phi\bar{\Xi}c(W_{k - 1,\bar{x}})},\nonumber\\
		& \frac{1 - 1/\eta_2}{\sqrt{24 T\sqrt{n_x}\phi\bar{\Xi} \bar{\Omega}}}, \frac{1 - 1/\eta_2}{\sqrt{24T\phi}}\bigg\}
	\end{align}
	where $c(W_{k - 1,\bar{x}}) := \Vert W_{k - 1,\bar{x}} \Vert \Vert W_{k - 1,\bar{x}}^\dag \Vert$ is the condition number of matrix $W_{k - 1,\bar{x}}$.
	Since $\underline{\lambda}_{W_{k - 1, \bar{x}}} \ge \hat{w}$, the condition number obeys $c(W_{k - 1,\bar{x}}) = \Vert W_{k - 1,\bar{x}} \Vert/\underline{\lambda}_{W_{k - 1, \bar{x}}} \ge 1$.
	This indicates that $\delta_{d,1}$ is independent of disturbance $d(k)$ and time $k$.
	This bound is similar to the bound in \cite[(37)]{depersis2020Low}.
	The only difference here is that $\Vert W_{k - 1}^\dag\Vert^{-1}$ is replaced by $\bar{w}$. This is because that the selected input sequence is $\bar{w}$-persistently exciting of order $n_x + 1$, and hence provides a positive constant lower bound on the smallest singular value of matrix $W_{k - 1}$.
	In conclusion, for sufficiently small disturbance $\Vert d(k)\Vert  \le \bar{d}$ with $\bar{d} < \delta_{d,1}$, a candidate solution of \eqref{eq:sdp_noise} can be constructed by $\eta_2(\bar{\gamma}(k), \bar{Q}(k),\bar{P}(k), \bar{L}(k), \bar{Q}(k)\bar{P}(k)^{-1}\bar{Q}(k)^\prime)$.
	Moreover, for any optimal solution of SDP \eqref{eq:sdp_noise}, i.e., $(\gamma^*(k), Q^*(k),P^*(k), L^*(k), V^*(k))$, matrix $K^*(k)$ satisfying $K^*(k) = U_{k - 1}Q^*(k)P^*(k)^{-1}$ is such that $A_i + B_iK^*(k)$ is Schur stable, which completes the proof.
\end{pf}

\subsection{Proof of Lemma \ref{lem:feasible_2}}
\label{sec:lem:feasible_2}
\begin{pf}
	Similar from the proof of Lemma \ref{lem:feasible_2} in \ref{pf:lem:feasible_2}, condition \eqref{eq:rankWk} holds due to the fact that input sequence ${u(k - T), \cdots, u(k - 1)}$ is $\bar{w}$-persistently exciting of order $n_x + 1$.
	\label{pf:lem:feasible_2}
	According to Assumption \eqref{as:T} and \eqref{as:dwell}, matrix $X_{k - 1}$ contains at least $N$ data from the same subsystem, it follows from \cite[Lemma 5]{Rotulo2021Online} that , SDP \eqref{eq:sdp_ideal} is feasible.
	In addition, it has been shown in the proof of Lemma \ref{lem:feasible_1} that under condition \eqref{eq:d_bound1}, a candidate solution of \eqref{eq:sdp_noise} can be constructed by $\eta_2(\bar{\gamma}(k), \bar{Q}(k),\bar{P}(k), \bar{L}(k), \bar{Q}(k)\bar{P}(k)^{-1}\bar{Q}(k)^\prime)$, where $(\bar{\gamma}(k), \bar{Q}(k), \bar{P}(k),\bar{L}(k))$ is any optimal solution of SDP \eqref{eq:sdp_ideal}.
	This completes the proof.
\end{pf}

\subsection{Proof of Lemma \ref{lem:rank}}
\label{sec:app:rank}
\begin{pf}
	For a given $\delta_{\epsilon}$ and $\epsilon(k) \in \mathbb{B}_{\delta_\epsilon}$, there exists a $\bar{w} >0$ such that sequence $\{\epsilon(k_j),\cdots, \epsilon(k_j + N - 1)\}$ is $\bar{w}$-persistently exciting for order $n_x + 1$.
	This indicates that ${u(k_j),\cdots, u(k_j + N - 1)}$ is $\bar{w}$-persistently exciting for order $n_x + 1$.
	One can further deduce that ${u(k-T),\cdots, u(k- 1)}$ is $\bar{w}$-persistently exciting for order $n_x + 1$ for all $k \in [k_j + N, k^j]$ with $j \in \mathbb{N}$.
	To be specific, let matrix $Y \in \mathbb{R}^{m \times n}$ with $n \ge m$ and ${\rm rank}(Y) = m$.
	Define matrix $Z = [Y,b_j] \in \mathbb{R}^{m \times (n + 1)}$ with $b_j \in \mathbb{R}^{m}$.
	Since $ZZ^{\prime}  = YY^{\prime} + b_jb_j^{\prime}$, the smallest singular value of matrices $Y$ and $Z$ satisfies $\underline{\lambda}_{Z} \ge \underline{\lambda}_{Y}$.
	Since the smallest singular value of matrix $H_{n_x + 1}(u_{[k_j, k_{j} + N - 1]})$ is lower bounded by $\bar{w}$, it can be deduced that the smallest singular value of matrix $H_{n_x + 1}(u_{[k_j, k_{j} + 2N-1]})$ is larger than or equal to $\bar{w}$.
	This implies that ${\rm rank}(W_{k - 1}) = n_x + n_u$ holds for all $k \in [k_j + N+1, k_j + 2N-1]$.
	According to Assumptions \ref{as:T} and the computational complexity concerns, $T = 2N - 1$.
	Based on the results in Lemmas \ref{lem:feasible_1} and \ref{lem:feasible_2}, we conclude that SDP \eqref{eq:sdp_noise} is feasible for all $k\in [k_j + N, k_j +T]$.
\end{pf}

\subsection{Proof of Lemma \ref{lem:relationship}}
\label{sec:app:lem:relationship}
\begin{pf}
	\label{pf:lem:relationship}
	Let $i \in \mathcal{M}$ denote the subsystem selected by $\sigma(k_{s_j})$, i.e., $i = \sigma(k), k \in [k_{s_j}, k_{s_{j+1}}]$.
	According to Lemma \ref{lem:feasible_1}, for $k \in [k_{s_j} + T, k_{s_{j+1}} - 1]$, the difference between the Lyapunov function at two consecutive time instants satisfies
	\begin{align*}
		\Delta \mathcal{W}_i(x(k+1)) &= \mathcal{W}_i(x(k + 1)) - \mathcal{W}_i(x(k))\\
		& = x(k)^{\prime}\!\mathcal{A}_i^{\prime}\!\cdot\! P_i(\star) -x(k)^{\prime}P_i(\star) \!+\! d(k)^{\prime}P_id(k)\nonumber\\
		&~~~+ \!2 x(k)^{\prime}\mathcal{A}_i^{\prime}P_id(k) \\
		&\le  \mathcal{W}^1(x(k))\!-\!(\beta_i/2)\Vert x(k)\Vert^2 \!+\! \bar{\lambda}_P\bar{d}^2\\
		&~~~+ 2\bar{\lambda}_P\Vert \mathcal{A}_i\Vert\Vert x(k)\Vert\bar{d} 
	\end{align*}
	where $\mathcal{W}^1(x(k)) = x(k)^{\prime}\!\mathcal{A}_i^{\prime}\!\cdot\! P_i(\star)-x(k)^{\prime}P_i x(k) + (\beta_i/2)\Vert x(k)\Vert^2$.
	Assuming that $k \notin \mathbb{I}_{\delta_V}$,  there exists a constant $\delta_x$ such that $\Vert x(k)\Vert \ge \delta_x$, which implies that
	\begin{subequations}
		\begin{align*}\label{eq:DeltaV}
			\Delta \mathcal{W}_i(x(k + 1))&\le \mathcal{W}^1(x(k))\\
			&~~~+ \!\Big(\! \underbrace{-\frac{\beta_i}{2}\! + \frac{\underline{\lambda}_P}{\delta_{x}^2}\bar{d}^2 + 2\frac{\underline{\lambda}_P}{\delta_{x}}\Vert \mathcal{A}_i\Vert\bar{d}}_{\Delta_{\mathcal{W}_{\bar{d}}}}\Big)\!\Vert x(k)\Vert^2.
		\end{align*}
	\end{subequations}
	Let $\bar{\beta} := \max_{i \in \mathcal{M}}\{\beta_i\}$ and  $\bar{\delta}_{\mathcal{A}} := \max_{i \in \mathcal{M}}\{\mathcal{A}_i \}$.
	If 
	$\bar{d} < \delta_{d,2}$ with 
	\begin{equation}\label{eq:d_bound2}
		\delta_{d,2} \!:=\min\Bigg\{\! \delta_{d,1},~ \delta_{x}\frac{-2\bar{\lambda}_{P}\bar{\delta}_{\mathcal{A}} \!+\! \sqrt{4 \bar{\lambda}_{P}^2\bar{\delta}_{\mathcal{A}}^2 \!+\! 2 \bar{\lambda}_{P}\bar{\beta}}}{2\bar{\lambda}_{P}}\Bigg\}
	\end{equation}
	and $\delta_{d,1}$ defined in \eqref{eq:d_bound1}, then $\Delta_{\mathcal{W}_{\bar{d}}} \le 0$ and $\Delta\mathcal{W}_i(x(k + 1)) \le \mathcal{W}^1(x(k))$.
	Leveraging \eqref{eq:boundV}, the following inequality holds for all $k \in [k_{s_j} + T, k_{s_{j+1}}]$ 
	\begin{equation}\label{eq:V:con}
		\mathcal{W}_i(x(k + 1)) \le \big(1 - \bar{\beta}/(2\bar{\lambda}_P)\big)\mathcal{W}_i(x(k)).
	\end{equation}
	
	In fact, due to disturbance $d(k)$, optimal solutions of SDP \eqref{eq:sdp_noise} for a subsystem $i \in \mathcal{M}$ activated at different times, i.e., $\sigma(k_{s_1}) = \sigma(k_{s_2})= \cdots = i$ with $k_{s_1} \ne k_{s_2} \ne \cdots$, are generally different.
	Suppose that condition \eqref{eq:rankWk} holds at $k = k_{s_j}, j \in \mathbb{N}$.
	According  to Lemma \ref{lem:feasible_1}, for a given $\eta_2 \ge 1$ and $\bar{d} < \delta_{d,2}$ with $\delta_{d,2}$ in \eqref{eq:d_bound2},  $\eta_2(\bar{\gamma}(k), \bar{Q}(k),\bar{P}(k), \bar{L}(k), \bar{Q}(k)\bar{P}(k)^{-1}\bar{Q}(k)^\prime)$ is a candidate solution of SDP \eqref{eq:sdp_noise}.
	Let $(\gamma^*(k), Q^*(k),P^*(k), L^*(k),\\ V^*(k))$ be an optimal solution of SDP \eqref{eq:sdp_noise}, and $K^*(k) = U_{k - 1}Q^*(k)P^*(k)^{-1}$.
	It follows from the last constraint of SDP \eqref{eq:sdp_noise} that $\bar{\lambda}_{P^*(k_{s_{j}})} \le {\rm tr}(P^*(k_{s_{j}})) \le\eta_2 \gamma^*(k_{s_{j}})  $.
	Since $(\bar{\gamma}(k), \bar{Q}(k), \bar{P}(k),\bar{L}(k))$ is the unique LQR solution of SDP \eqref{eq:sdp_ideal} \cite[Lemma 4]{Rotulo2021Online}, it holds that $\bar{\gamma}(k_{s_j}) = \bar{\gamma}_i$ for all $j \in \mathbb{N}$.
	Let 
	\begin{equation}
		\label{eq:check:lambda}
		\check{\lambda}_0 := 1 - \frac{\bar{\beta}}{2\max_{i \in \mathcal{M}} \eta_2 \bar{\gamma}_i}.
	\end{equation}
	Hence, for any $i \in \mathcal{M}$ and $j \in \mathbb{N}$, matrix $K^*(k_{s_j})$ is such that $A_i+ B_iK^*(k_{s_j})$ is Schur stable, the Lyapunov function converges following $\mathcal{W}_i(x(k + 1)) \le \check{\lambda}_0\mathcal{W}_i(x(k))$ for all $k \in [k_{s_j} + T, k_{s_{j+1}}]$.
	This further implies that for some $\lambda_0 \in [\check{\lambda}_0, 1)$, it holds that $\mathcal{V}(x(k+ 1)) \le \check{\lambda}_0\mathcal{V}(x(k))$.

	On the other hand, for $k = k^j$, it follows from \eqref{eq:uk}--\eqref{eq:P_strategy} that  $\mathcal{V}(x(k^j+1)) \le \lambda_0 \mathcal{V}(x(k^j))$, $P(k^j+1) = P(k^j)$, and $K(k^j+1) = K(k^j)$.
The difference between the auxiliary function at two consecutive time instants satisfies
\begin{align*}
	\Delta \mathcal{V}(x(k^j+1)) &= \mathcal{V}(x(k^j+1)) - \mathcal{V}(x(k^j))\\
	& = x(k^j)^{\prime}\!\mathcal{A}(k^j\!+\!1)^{\prime}\!\cdot\! P(k^j)(\star) \\
	&~~~ -x(k^j)^{\prime}P(k^j)(\star) \!+\! d(k^j)^{\prime}P(k^j)d(k^j)\nonumber\\
	&~~~+ \!2 x(k^j)^{\prime}\mathcal{A}(k^j\!+\!1)^{\prime}P(k^j)d(k^j) \\
	&= \mathcal{V}_1(x(k^j))\!-\!(\bar{\beta}/2)\Vert x(k^j)\Vert^2 \!+\! \bar{\lambda}_{P(k^j)}\bar{d}^2\\
	&~~~+ 2\bar{\lambda}_{P(k^j)}\Vert \mathcal{A}(k^j\!+\!1)\Vert\Vert x(k^j)\Vert\bar{d} \\
	&\le \mathcal{V}_1(x(k^j))+ \Big( \underbrace{-\frac{\bar{\beta}}{2}+ \frac{\underline{\lambda}_{P(k^j)}}{\delta_{x}^2}\bar{d}^2} \\
	&~~~ \underbrace{+ 2\frac{\underline{\lambda}_{P(k^j)}}{\delta_{x}}\Vert \mathcal{A}(k^j\!+\!1)\Vert\bar{d}}_{\Delta_{\mathcal{V}_{\bar{d}}}}\Big)\!\Vert x(k^j)\Vert^2
\end{align*}
where $\mathcal{A}(k^j\!+\!1) \!:=\! A_{\sigma(k^j\!+\!1)} + B_{\sigma(k^j\!+\!1)}K(k^j)$, $\mathcal{V}_1(x(k^j\!+\!1)) = x(k^j)^{\prime}\!\mathcal{A}(k^j\!+\!1)^{\prime}\!\cdot\! P(k^j)(\star)-x(k^j)^{\prime}P(k^j)x(k^j) + \bar{\beta}/2\Vert x(k^j)\Vert^2$, and the last inequality holds since $\mathcal{V}(x(k^j)) \ge \delta_V$.
Since $\bar{d} < \delta_{d,2}$, it follows that $\Delta_{\mathcal{V}_{\bar{d}}}\le 0$ and $\Delta \mathcal{V}(x(k^j + 1))\le \mathcal{V}_1(x(k^j))$.
Since $\lambda_0 \in [\check{\lambda}_0, 1)$ with $\check{\lambda}_0$ in \eqref{eq:check:lambda}, one has that
\begin{align*}
	\Delta\mathcal{V}(x(k^j + 1)) &= \mathcal{V}(x(k^j+1)) - \mathcal{V}(x(k^j)) \\
	&\le (\lambda_0 - 1)\mathcal{V}(x(k^j))\le - (\bar{\beta}/2)\Vert x(k^j)\Vert^2
\end{align*}
hence $x(k^j)^{\prime}\!\mathcal{A}(k^j\!+\!1)^{\prime}\!\cdot\! P(k^j)(\star)-x(k^j)^{\prime}P(k^j)x(k^j) + {\bar{\beta}}/{2}\Vert x(k^j)\Vert^2 \le - {\bar{\beta}}/{2}\Vert x(k^j)\Vert^2$ and consequently $\mathcal{A}(k^j\!+\!1)^{\prime} P(k^j)\mathcal{A}(k^j\!+\!1)- P(k^j)\le -\bar{\beta}I$.
This implies that $K(k^j)$ derived from \eqref{eq:sdp_noise} stabilizes subsystem $(A_{\sigma(k^j + 1)}, B_{\sigma(k^j + 1)})$.
Hence, it follows from Lemma \ref{lem:feasible_1} that $k^j \in [k_{s_j} + T, k_{s_{j+1}}]$.

Moreover, there exists some constant $\hat{\lambda}_0$ obeying $\check{\lambda}_0 <\hat{\lambda}_0 \le 1$ such that condition $\mathcal{V}(x(k^j - 1)) > \lambda_0 \mathcal{V}(x(k^j - 2))$ holds when $K(k^j - 2)$ cannot stabilize $(A_{\sigma(k^j - 1)}, B_{\sigma(k^j - 1)})$ or the Lyapunov function converges with a rate smaller than $\check{\lambda}_0$.
Since the disturbance satisfies \eqref{eq:d_bound2}, based on Lemma \ref{lem:feasible_2}, this only occurs when matrices $X_{k - 1}$ and $X_{k}$ contain samples generated from two subsystems. 
Hence, $k^j - 1 \in [k_{s_j} , k_{s_j} + T - 1]$, and consequently $k^j = k_{s_j} + T$ holds  for all $j \in \mathbb{N}$.

Similarly, for time $k_j$, condition $\mathcal{V}(x(k_j)) > \lambda_0 \mathcal{V}(x(x_j - 1))$ means that $k_j \in [k_{s_j}, k_{s} + T - 1]$, and condition $\mathcal{V}(x(k_j-1)) \le \lambda_0 \mathcal{V}(x(x_j - 2))$ implies that $k_j - 1 \in [k_{s-1}  + T, k_{s}]$.
Hence, $k_j = k_{s_j}$ for all $j \in \mathbb{N}$.
This further implies that $k_j$ and $k^j$ in sequences $\{k_j \}_{j \in \mathbb{N}}$ and $\{k^j \}_{i \in \mathbb{N}}$ are ordered such that $0=k_0<k_0 + N< k^0< k_1  <k_1 + N < k^1 < \cdots$.

Noticing that system \eqref{eq:sys} may switch its mode at $k_{s_j}$ while condition $\mathcal{V}(x(k_{s_j})) \le \lambda_0 \mathcal{V}(x(k_{s_j} - 1))$ still holds.
According to \eqref{eq:uk}--\eqref{eq:P_strategy}, SDP \eqref{eq:sdp_noise} is not solved, and both matrices $K(k_{s_j})$ and $P(k_{s_j})$ remain unchanged.
In this case, the system does not aware that a switching happens, and the current controller can guarantee that the convergence of Lyapunov function satisfies \eqref{eq:V:con}.
Hence, such switches will not affect system stability.  
\end{pf}

				\bibliographystyle{plainnat}
				\bibliography{bible3}

\begin{thebibliography}{37}
\providecommand{\natexlab}[1]{#1}
\providecommand{\url}[1]{\texttt{#1}}
\expandafter\ifx\csname urlstyle\endcsname\relax
  \providecommand{\doi}[1]{doi: #1}\else
  \providecommand{\doi}{doi: \begingroup \urlstyle{rm}\Url}\fi

\bibitem[{\AA}str{\"o}m and Wittenmark(1989)]{aastrom2013adaptive}
K.~J. {\AA}str{\"o}m and B.~Wittenmark.
\newblock \emph{Adaptive {C}ontrol}.
\newblock Addison-Wesley, MA, USA, 1989.

\bibitem[Baggio et~al.(2021)Baggio, Bassett, and Pasqualetti]{baggio2020Data}
G.~Baggio, D.~S. Bassett, and F.~Pasqualetti.
\newblock Data-driven control of complex networks.
\newblock \emph{Nat. Commun.}, \penalty0 (1429):\penalty0 1--13, Mar. 2021.

\bibitem[Berberich et~al.(2021)Berberich, K{\"o}hler, M{\"u}ller, and
  Allg{\"o}wer]{berberich2019data}
J.~Berberich, Johannes K{\"o}hler, Matthias~A M{\"u}ller, and Frank
  Allg{\"o}wer.
\newblock Data-driven model predictive control with stability and robustness
  guarantees.
\newblock \emph{IEEE Trans. Autom. Control}, 66\penalty0 (4):\penalty0
  1702--1717, Jun. 2021.

\bibitem[Bianchi et~al.(2022)Bianchi, Grammatico, and
  Cort{\'e}s]{bianchi2022data}
M.~Bianchi, S.~Grammatico, and J.~Cort{\'e}s.
\newblock Data-driven stabilization of switched and constrained linear systems.
\newblock \emph{arXiv:2208.11392}, Aug. 2022.

\bibitem[Cardim et~al.(2009)Cardim, Teixeira, Assuncao, and
  Covacic]{Cardim2009variable}
R.~Cardim, M.~C.~M. Teixeira, E.~Assuncao, and M.~R. Covacic.
\newblock Variable-structure control design of switched systems with an
  application to a {DC}–{DC} power converter.
\newblock \emph{IEEE Trans. Ind. Electron.}, 56\penalty0 (9):\penalty0
  3505--3513, Jul. 2009.

\bibitem[Chua et~al.(2018)Chua, Calandra, McAllister, and Levine]{Chua2018deep}
K.~Chua, R.~Calandra, R.~McAllister, and S.~Levine.
\newblock Deep reinforcement learning in a handful of trials using
  probabilistic dynamics models.
\newblock In \emph{Proc. of Adv. Neural Inform. Process. Syst.}, pages
  4759--4770, Montréal Canada, Dec. 3-8, 2018.

\bibitem[{Coulson} et~al.(2019){Coulson}, {Lygeros}, and
  {Dörfler}]{Coulson2019cdc}
J.~{Coulson}, J.~{Lygeros}, and F.~{Dörfler}.
\newblock Regularized and distributionally robust data-enabled predictive
  control.
\newblock In \emph{Proc. of IEEE Conf. Decis. Control}, pages 2696--2701, Nice,
  France, Dec. 11-13, 2019.

\bibitem[Coulson et~al.(2022)Coulson, van Waarde, and
  D{\"o}rfler]{coulson2022robust}
J.~Coulson, H.~van Waarde, and F.~D{\"o}rfler.
\newblock Robust fundamental lemma for data-driven control.
\newblock \emph{arXiv:2205.06636}, May, 2022.

\bibitem[{De Persis} and {Tesi}(2020)]{persis2020data}
C.~{De Persis} and P.~{Tesi}.
\newblock Formulas for data-driven control: {S}tabilization, optimality, and
  robustness.
\newblock \emph{IEEE Trans. Autom. Control}, 65\penalty0 (3):\penalty0
  909--924, Mar. 2020.

\bibitem[{De Persis} and {Tesi}(2021)]{depersis2020Low}
C.~{De Persis} and P.~{Tesi}.
\newblock Low-complexity learning of linear quadratic regulators from noisy
  data.
\newblock \emph{Automatica}, 128:\penalty0 109548, Jun. 2021.

\bibitem[De~Persis et~al.(2022)De~Persis, Postoyan, and Tesi]{de2022event}
C.~De~Persis, R.~Postoyan, and P.~Tesi.
\newblock Event-triggered control from data.
\newblock \emph{arXiv:2208.11634}, Aug. 2022.

\bibitem[{Dörfler} et~al.(2023){Dörfler}, {Coulson}, and
  {Markovsky}]{Coulson2021Bridging}
F.~{Dörfler}, J.~{Coulson}, and I.~{Markovsky}.
\newblock Bridging direct \& indirect data-driven control formulations via
  regularizations and relaxations.
\newblock \emph{IEEE Trans. Autom. Control}, 68\penalty0 (2):\penalty0
  883--897, Feb. 2023.

\bibitem[Eising et~al.(2022)Eising, Liu, Mart{\'\i}nez, and
  Cort{\'e}s]{eising2022using}
J.~Eising, S.~Liu, S.~Mart{\'\i}nez, and J.~Cort{\'e}s.
\newblock Using data informativity for online stabilization of unknown switched
  linear systems.
\newblock In \emph{Proc. of IEEE Conf. Decis. Control}, pages 8--13, Cancun,
  Mexico, Dec. 06-09, 2022.

\bibitem[Guo et~al.(2022)Guo, De~Persis, and Tesi]{guo2022data}
M.~Guo, C.~De~Persis, and P.~Tesi.
\newblock Data-driven stabilizer design and closed-loop analysis of general
  nonlinear systems via taylor's expansion.
\newblock \emph{arXiv:2209.01071}, Apr. 2022.

\bibitem[Hjalmarsson et~al.(1998)Hjalmarsson, Gevers, Gunnarsson, and
  Lequin]{hjalmarsson1998iterative}
H.~Hjalmarsson, M.~Gevers, S.~Gunnarsson, and O.~Lequin.
\newblock Iterative feedback tuning: {T}heory and applications.
\newblock \emph{IEEE Control Syst. Mag.}, 18\penalty0 (4):\penalty0 26--41,
  Aug. 1998.

\bibitem[Hou and Wang(2013)]{Hou2013from}
Z.~Hou and Z.~Wang.
\newblock From model-based control to data-driven control: {S}urvey,
  classification and perspective.
\newblock \emph{Inf. Sci.}, 235:\penalty0 3--35, Jun. 2013.

\bibitem[Hu et~al.(2023)Hu, De~Persis, and Tesi]{hu2023learning}
Z.~Hu, C.~De~Persis, and P.~Tesi.
\newblock Learning controllers from data via kernel-based interpolation.
\newblock \emph{arXiv:2304.09577}, Apr. 2023.

\bibitem[{Jiang} et~al.(1994){Jiang}, {Teel}, and {Praly}]{Z1994Small}
Z.~{Jiang}, A.~R. {Teel}, and L.~{Praly}.
\newblock Small-gain theorem for {ISS} systems and applications.
\newblock \emph{Math. Control Signals Syst.}, 7:\penalty0 95--120, Mar. 1994.

\bibitem[Kang and You(2023)]{kang2023minimum}
S.~Kang and K.~You.
\newblock Minimum input design for direct data-driven property identification
  of unknown linear systems.
\newblock \emph{Automatica}, 156:\penalty0 111130, Oct. 2023.

\bibitem[Krishnan and Pasqualetti(2021)]{krishnan2021On}
V.~Krishnan and F.~Pasqualetti.
\newblock On direct vs indirect data-driven predictive control.
\newblock In \emph{Proc. of IEEE Conf. Decis. Control}, pages 736--741, Austin,
  TX, USA, Dec. 14-17, 2021.

\bibitem[Lee and Jiang(2008)]{lee2008uniform}
T.C. Lee and Z.~Jiang.
\newblock Uniform asymptotic stability of nonlinear switched systems with an
  application to mobile robots.
\newblock \emph{IEEE Trans. Autom. Control}, 53\penalty0 (5):\penalty0
  1235--1252, Aug. 2008.

\bibitem[Li et~al.(2023{\natexlab{a}})Li, De~Persis, Tesi, and
  Monshizadeh]{li2023data}
L.~Li, C.~De~Persis, P.~Tesi, and N.~Monshizadeh.
\newblock Data-based transfer stabilization in linear systems.
\newblock \emph{IEEE Trans. Autom. Control}, Nov. 2023{\natexlab{a}}.
\newblock \doi{10.1109/TAC.2023.3330792}.

\bibitem[Li et~al.(2023{\natexlab{b}})Li, Wang, Sun, Wang, and
  Chen]{li2022robust}
Y.~Li, X.~Wang, J.~Sun, G.~Wang, and J.~Chen.
\newblock Data-driven consensus control of fully distributed event-triggered
  multi-agent systems.
\newblock \emph{Sci. {CHINA} Inf. Sci.}, 66\penalty0 (5):\penalty0 152202--,
  May, 2023{\natexlab{b}}.

\bibitem[Liu et~al.(2017)Liu, Zhang, Shi, and Zhao]{Liu2017sliding}
M.~Liu, L.~Zhang, P.~Shi, and Y.~Zhao.
\newblock Sliding mode control of continuous-time {M}arkovian jump systems with
  digital data transmission.
\newblock \emph{Automatica}, 80:\penalty0 200--209, Jun. 2017.

\bibitem[Liu et~al.(2023)Liu, Sun, Wang, Bullo, and Chen]{Liu2021data}
W.~Liu, J.~Sun, G.~Wang, F.~Bullo, and J.~Chen.
\newblock Data-driven resilient predictive control under denial-of-service.
\newblock \emph{IEEE Trans. Autom. Control}, 68\penalty0 (8):\penalty0
  4722--4737, Aug. 2023.

\bibitem[Mhaskar et~al.(2005)Mhaskar, El-Farra, and
  Christofides]{Mhaskar2005predictive}
P.~Mhaskar, N.H. El-Farra, and P.D. Christofides.
\newblock Predictive control of switched nonlinear systems with scheduled mode
  transitions.
\newblock \emph{IEEE Trans. Autom. Control}, 50\penalty0 (11):\penalty0
  1670--1680, Nov. 2005.

\bibitem[Rotulo et~al.(2022)Rotulo, {De Persis}, and Tesi]{Rotulo2021Online}
M.~Rotulo, C.~{De Persis}, and P.~Tesi.
\newblock Online learning of data-driven controllers for unknown switched
  linear systems.
\newblock \emph{Automatica}, 145:\penalty0 110519, Nov. 2022.

\bibitem[Rueda-Escobedo et~al.(2022)Rueda-Escobedo, Fridman, and
  Schiffer]{rueda2020data}
J.~G. Rueda-Escobedo, E.~Fridman, and J.~Schiffer.
\newblock Data-driven control for linear discrete-time delay systems.
\newblock \emph{IEEE Trans. Autom. Control}, 67\penalty0 (7):\penalty0
  3321--3336, Jul. 2022.

\bibitem[Sassano and Astolfi(2020)]{sassano2020combining}
M.~Sassano and A.~Astolfi.
\newblock Combining {P}ontryagin's principle and dynamic programming for linear
  and nonlinear systems.
\newblock \emph{IEEE Trans. Autom. Control}, 65\penalty0 (12):\penalty0
  5312--5327, Sept. 2020.

\bibitem[van Waarde et~al.(2020)van Waarde, Camlibel, and
  Mesbahi]{van2020noisy}
H.~J. van Waarde, M.~Kanat Camlibel, and Mehran Mesbahi.
\newblock From noisy data to feedback controllers: non-conservative design via
  a matrix {S}-lemma.
\newblock \emph{IEEE Trans. Autom. Control}, 67\penalty0 (1):\penalty0
  162--175, Jan. 2020.

\bibitem[{van Waarde} et~al.(2020){van Waarde}, {Eising}, {Trentelman}, and
  {Camlibel}]{van2020data}
H.~J. {van Waarde}, J.~{Eising}, H.~L. {Trentelman}, and M.~K. {Camlibel}.
\newblock Data informativity: {A} new perspective on data-driven analysis and
  control.
\newblock \emph{IEEE Trans. Autom. Control}, 65\penalty0 (11):\penalty0
  4753--4768, Jan. 2020.

\bibitem[Wang et~al.(2023)Wang, Sun, Berberich, Wang, Allgöwer, and
  Chen]{Wang2021delay}
X.~Wang, J.~Sun, J.~Berberich, G.~Wang, Frank Allgöwer, and J.~Chen.
\newblock Data-driven control of dynamic event-triggered systems with delays.
\newblock \emph{Int. J. Robust Nonlin.}, 33\penalty0 (12):\penalty0 7071--7093,
  May, 2023.

\bibitem[Wang et~al.(2021)Wang, Sun, and Chen]{finite2021wang}
Z.~Wang, J.~Sun, and J.~Chen.
\newblock Finite-time integral input-to-state stability for switched nonlinear
  time-delay systems with asynchronous switching.
\newblock \emph{Int. J. Robust Nonlin.}, 31\penalty0 (9):\penalty0 3929--3954,
  Mar. 2021.

\bibitem[{Willems} et~al.(2005){Willems}, {Markovsky}, {Rapisarda}, and {De
  Moor}]{willems2005note}
J.~C. {Willems}, I.~{Markovsky}, P.~{Rapisarda}, and B.~L.~M. {De Moor}.
\newblock A note on persistency of excitation.
\newblock \emph{Syst. Control Lett.}, 56\penalty0 (4):\penalty0 325--329, May,
  2005.

\bibitem[Wu and Meng(2023)]{wu2023data}
Y.~Wu and D.~Meng.
\newblock Data-based trackability criteria and control design for disturbed
  learning systems.
\newblock \emph{Automatica}, 155:\penalty0 111113, Sept. 2023.

\bibitem[X.{ Wang} et~al.(2023)X.{ Wang}, {Berberich}, {Sun}, {Wang},
  {Allg{\"o}wer}, and {Chen}]{Wang2021data}
X.{ Wang}, J.~{Berberich}, J.~{Sun}, G.~{Wang}, F.~{Allg{\"o}wer}, and
  J.~{Chen}.
\newblock Data-driven control of event- and self-triggered discrete-time
  systems.
\newblock \emph{IEEE Trans. Cybern.}, 53\penalty0 (9):\penalty0 6066--6079,
  Sept. 2023.

\bibitem[Zhao et~al.(2023)Zhao, D{\"o}rfler, and You]{zhao2023data}
F.~Zhao, F.~D{\"o}rfler, and K.~You.
\newblock Data-enabled policy optimization for the linear quadratic regulator.
\newblock \emph{arXiv:2303.17958}, Sept. 2023.

\end{thebibliography}

			\end{document}